\documentclass[%
 reprint,
 amsmath,amssymb,
 aps,
prb,
]{revtex4-2}

\usepackage{xr-hyper}
\usepackage{hyperref}
\usepackage{xcite}
\usepackage{braket}
\usepackage{xcolor}
\usepackage{amsmath}
\usepackage{siunitx}
\usepackage{multirow}

\usepackage{graphicx}
\usepackage{dcolumn}
\usepackage{bm}

\begin{document}


\title{A quantum model for rf-SQUIDs based metamaterials enabling 3WM and 4WM Traveling Wave Parametric Amplification}

\author{Angelo Greco}
\affiliation{INRiM, Istituto Nazionale di Ricerca Metrologica, Strada delle Cacce 91, 10135 Torino, Italy}
\affiliation{Department of Electronics and Telecommunications, PoliTo, Corso Castelfidardo 39, 10129 Torino, Italy}
\author{Luca Fasolo}
\affiliation{INRiM, Istituto Nazionale di Ricerca Metrologica, Strada delle Cacce 91, 10135 Torino, Italy}
\affiliation{Department of Electronics and Telecommunications, PoliTo, Corso Castelfidardo 39, 10129 Torino, Italy}
\author{Alice Meda, Luca Callegaro}
\affiliation{INRiM, Istituto Nazionale di Ricerca Metrologica, Strada delle Cacce 91, 10135 Torino, Italy}
\author{Emanuele Enrico}
    \email{Corresponding author: e.enrico@inrim.it}
\affiliation{INRiM, Istituto Nazionale di Ricerca Metrologica, Strada delle Cacce 91, 10135 Torino, Italy}
\affiliation{INFN, Trento   Institute for  Fundamental  Physics  and  Applications,  I-38123,  Povo,  Trento,  Italy}

\date{\today}

\begin{abstract}
A quantum model for Josephson-based metamaterials working in the Three-Wave Mixing (3WM) and Four-Wave Mixing (4WM) regimes at the single-photon level is presented. The transmission line taken into account, namely Josephson Traveling Wave Parametric Amplifier (JTWPA), is a bipole composed of a chain of rf-SQUIDs which can be biased by a DC current or a magnetic field to activate the 3WM or 4WM nonlinearities. The model exploits a Hamiltonian approach to analytically determine the time evolution of the system both in the Heisenberg and interaction pictures. The former returns the analytic form of the gain of the amplifier, while the latter allows recovering the probability distributions vs. time of the photonic populations, for multimodal Fock and coherent input states. The dependence of the metamaterial’s nonlinearities is presented in terms of circuit parameters in a lumped model framework while evaluating the effects of the experimental conditions on the model validity.
\end{abstract}

\maketitle

\section{Introduction}

Superconducting amplifiers are nowadays widely used for the manipulation of single photons in several ranges of the electromagnetic spectrum. From microwaves to X-rays these devices have shown unrivalled performances for what concerns quantum efficiency, resolving power and added noise, compared to their solid-state counterparts \cite{Hadfield2009,Natarajan2012,Ullom2014,Fukuda2011,Miller2003,Lita2008}.
The peculiar characteristics of superconducting materials allow engineering highly performing resonators and cavities, characterised by a quality factor of the order of $\approx 10^{10}$ \cite{Yang2020,Malnou2018,Castellanos-Beltran2009,Eichler2014}. Indeed,  resonator-based superconducting amplifiers show a quite high gain, in the range of $\SI{20}{\decibel}$ \cite{Castellanos-Beltran2009}; however, they are subjected to limited bandwidth making them unsuitable for the multiplexing required in complex systems.\\
Josephson Traveling Waves Parametric Amplifiers (JTWPAs) and Kinetic Inductance Traveling Wave Amplifiers (KITs) promise to be appropriate devices for this aim in the microwave regime, showing in principle valuable multiplexing capabilities due to their wide bandwidth \cite{White2015}. Indeed, it has been shown how the Four-Wave Mixing (4WM) induced in all the Kerr-like media allows amplifying very tiny signals over several $\si{\giga\hertz}$ bandwidths with a nearly quantum-limited noise \cite{White2015,Macklin2015,Zobrist2019,Chaudhuri2017}.
Nevertheless, recent papers show that enabling the Three-Wave Mixing (3WM) interaction, through the introduction of a quadratic nonlinearity in the medium, could provide several benefits and experimental simplifications for what concerns feasibility and integration capabilities. In particular, a three-wave mixer generally requires a lower input pump power, easier output filtering and shows a higher dynamic range  \cite{Vissers2016,Zorin2016,Zorin2017}.
These distinctive characteristics make JTWPAs working in 3WM excellent candidates for the readout of quantum-limited detectors (e.g., rf-SETs, rf-SQUIDs), by preserving the quantum properties of their outputs \cite{Buehler2005,Aassime2001,Shawn2018}. Moreover, a three-wave mixer can be a promising candidate for the generation of heralded photons pairs, since it naturally enables Parametric Down Conversion (PDC) \cite{Guo2017}.\\
In this framework, we develop a quantum model, based on previous theoretical works \cite{vanderreep, Grimsmo2017}, for a recently proposed JTWPA concept \cite{Zorin2016} covering both the 3WM and 4WM regimes. Previous classical descriptions in terms of electromagnetic waves \cite{Zorin2016,Zorin2017} were limited to the high power range, completely neglecting any description of the light-matter interaction at the single photon level. Our theory exploits circuit-QED techniques to model a JTWPA made up of a chain of rf-SQUIDs capacitively shunted to ground. The proposed layout can be biased by a DC current or an externally applied magnetic field to activate 3WM or 4WM of the microwave traveling modes. The quantum description allows to analytically treat important figures of merit of the amplifier as the gain, squeezing and the time evolution of arbitrary quantum states at the single-photon level.\\
The main results of the paper are reported in Section \ref{Results}. In particular, Subsection \ref{ch:RF-SQUIDs array embedded in a transmission line} reports the Hamiltonian in first quantization formalism, based on the circuit model of a nonlinear lossless transmission line. Then, Subsection \ref{ch:Second quantization framework} is dedicated to developing the theory through the occupation number formalism, and a 3WM/4WM Hamiltonian is found. A selection of modes follows in Subsection \ref{sec:PDC}, leading to model the 3WM/4WM quantum mechanical phenomena in the Heisenberg picture. Solving the dynamics of the system (i.e., Langevin equations) allows to analytically calculate the gain, noise figures and squeezing capabilities of the amplifier. In Subsection \ref{interaction_of_quantum_states} the time evolution of Fock and coherent input states due to nonlinear interactions is analytically treated and on these bases various examples of photon statistics in the Fock space are calculated.

\section{Results}
\label{Results}

\subsection{\label{ch:RF-SQUIDs array embedded in a transmission line}Hamiltonian of a rf-SQUIDs array embedded in a transmission line}

\begin{figure}
    \centering
    \includegraphics[width=\linewidth]{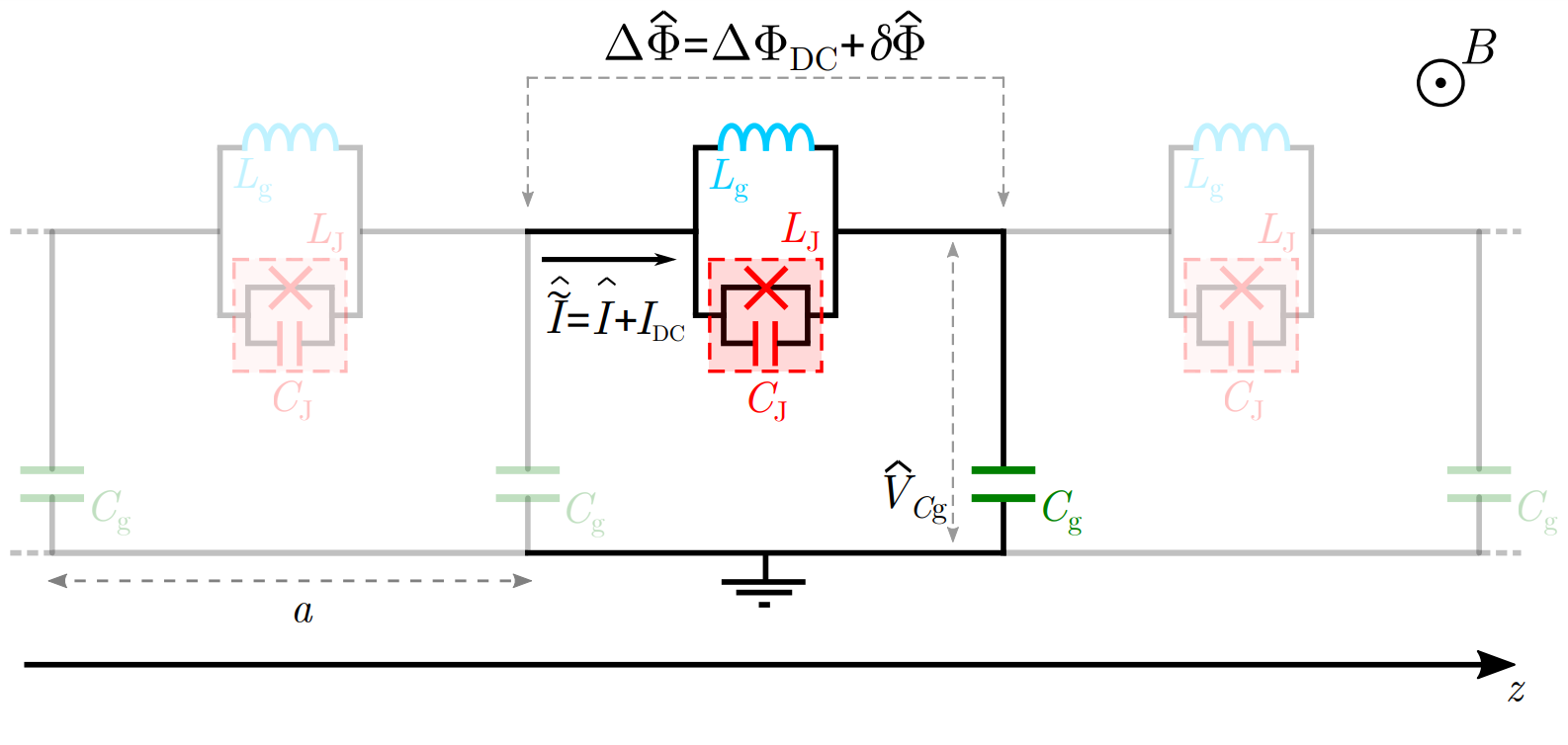}
    \caption{Electrical equivalent of a repetition of three elementary cells (periodicity $a$) of the rf-SQUIDs based JTWPA. Each cell consists of a superconducting loop containing a geometrical inductance $L_\mathrm{g}$, a Josephson junction, with an associated capacitance $C_\mathrm{J}$ and inductance $L_\mathrm{J}$, and a ground capacitor $C_\mathrm{g}$. The series can be biased both through an external DC magnetic field $B$ and a flowing current $I_{\mathrm{DC}}$. $\Delta\hat{\Phi}$ is the magnetic flux difference across the nodes of a cell, while $\hat{V}_{C_\mathrm{g}}$ is the voltage drop across the ground capacitor.}
    \label{fig:RfSquid_Array}
\end{figure}

The JTWPA recently proposed \cite{Zorin2016} and theoretically quantum mechanically treated in this paper can be modelled as an array of
rf-SQUIDs embedded in a superconducting transmission line. In the following, the Hamiltonian of the system will be derived as a function of its circuit parameters. As represented in Figure\;\ref{fig:RfSquid_Array}, each elementary cell is composed by a superconducting loop containing a Josephson junction (with its associated capacitance $C_\mathrm{J}$ and inductance $L_\mathrm{J}$) and a geometrical inductance $L_\mathrm{g}$. Furthermore, each loop is coupled to ground through a capacitor $C_\mathrm{g}$. The system taken into account is non-dissipative and, for the sake of simplicity, all the elementary cells are considered identical. The length of the elementary cell along the $z$-direction (i.e., the propagating direction of the modes) is defined as $a$.\\
In presence of an electromagnetic field, each of these cells stores a certain amount of energy that can be expressed as a function of the conjugate coordinates $\hat{\Phi}$ and $\hat{Q}$, the generalized magnetic flux and charge at a certain node respectively, obeying to the commutation relation $[\hat\Phi,\hat Q]=i\hbar$. The total amount of energy can be computed as the sum of the energy stored in each of its components (see Supplemental \ref{appendix:hamiltonian}). 
Moreover, being the system under analysis a repetition of identical elementary units, the total energy stored in the whole medium can be expressed as the sum of the energy stored in each cell.\\
Under the assumption that the differences between the $\hat{\Phi}$ (and $\hat{Q}$) of a couple of consecutive nodes are small enough, these operators can be considered as functions of both time and space (i.e., $\hat{\Phi}(z,t)$ and $\hat{Q}(z,t)$). 
We then define the flux difference between two subsequent nodes as
\begin{align}
    \Delta \hat\Phi (z,t)=\hat\Phi(z+a,t)-\hat\Phi(z,t)
\end{align}
We can define the Hamiltonian of the system like the sum of the energies stored in every single cell of the device. The circuit elements which appear are discrete, so every cell has its own ground capacitor, Josehpson capacitor, geometrical inductance and Josephson junction. The sum runs over the index $n$ which labels all the cells.
\begin{widetext}
\begin{subequations}
\begin{eqnarray}
\hat{H} &=& \sum_{n=1}^N \hat{H}_\mathrm{n}
\\
 &=& \sum_{n=1}^N \left( \hat{H}_\mathrm{{L_g}} + \hat{H}_\mathrm{{L_j}} + \hat{H}_\mathrm{{C_j}} + \hat{H}_\mathrm{{C_g}} \right)
 \label{eq:linear_density_H}
\\
 &=& \sum_{n=1}^N \left( \frac{1}{2L_\mathrm{g}}\Delta\hat{\Phi}(na,t)^2 + \varphi_0I_\mathrm{c}\left(1-\cos{\left(\frac{\Delta \hat{\Phi}(na,t)}{\varphi_0}\right)}\right) + \frac{C_\mathrm{J}}{2}\left(\frac{\partial \Delta\hat{\Phi}(na,t)}{\partial t}\right)^2 + \frac{1}{2C_\mathrm{g}}\hat{Q}(na,t)^2 \right) \label{eq:linear_density_H_2}
\end{eqnarray}
\end{subequations}
\end{widetext}

where in the right-hand side of equation \eqref{eq:linear_density_H} one can recognize respectively the energy associated to the geometrical inductance $L_\mathrm{g}$, the Josephson inductance $L_\mathrm{J}$, the Josephson capacitance $C_\mathrm{J}$ and the ground capacitance $C_\mathrm{g}$. $N$ is the number of unit cells composing the transmission line, $I_\mathrm{c}$ is the critical current of the Josephson junction and $\varphi_0 =\Phi_0/2\pi= \hbar/2e$ is the reduced flux quantum.\\
As can be seen, the flux difference function $\Delta\hat\Phi(z,t)$ is defined for every $z$ but is calculated at discrete points in correspondence to the multiple integers $an$ of the unit cell length $a$.
In order to compute the Hamiltonian we can switch from a discrete sum to an approximated continuous sum substituting the summation sign with an integral \cite{quantumlight}, adding the scale factor $a$ (length of a unit cell), which turns the discrete components in components per unit length. The integration upper limit is the length of the amplifier $l=Na$, where $N$ is the number of unit cells.
\begin{widetext}
\begin{align}\label{eq:first_quantization_hamiltonian}
    \hat{H} &= \int_{0}^l \left( \frac{1}{2L_\mathrm{g}}\Delta\hat{\Phi}(z,t)^2 + \varphi_0I_\mathrm{c}\left(1-\cos{\left(\frac{\Delta \hat{\Phi}(z,t)}{\varphi_0}\right)}\right) + \frac{C_\mathrm{J}}{2}\left(\frac{\partial \Delta\hat{\Phi}(z,t)}{\partial t}\right)^2 + \frac{1}{2C_\mathrm{g}}\hat{Q}(z,t)^2 \right)\frac{dz}{a}
\end{align}
\end{widetext}
The integral can be considered as an approximation of the discrete sum in Equation \eqref{eq:linear_density_H_2}, where the flux difference across two subsequent cells is to be considered due to the presence of a finite number of Josephson junctions, for which the energy is defined in relation with the flux difference across them.\\
The presence of an external magnetic field or a DC current through the line induces a constant component in the flux difference across a cell. This means that $\Delta \hat \Phi(z,t)$ can be considered as the sum of two components, a constant one $\Delta {\Phi}_{\mathrm{DC}}$ and a time-dependent one $\delta\hat{\Phi}(z,t)$
\begin{equation}\label{eq:2}
    \Delta\hat{\Phi}(z,t) = \Delta{\Phi}_{\mathrm{DC}} + \delta\hat{\Phi}(z,t)
\end{equation}

\subsection{\label{ch:Second quantization framework}Second quantization framework}

Here the Hamiltonian will be expressed in terms of ladder operators. In this view, the voltage drop on the ground capacitors $C_\mathrm{g}$ can be expressed using a mode decomposition assuming that sinusoidal waves are passing through the line \citep{Vool2017b}
\begin{equation} \label{eq9}
    \hat V_{C_\mathrm{g}}(z,t)=\sum_n \sqrt{\frac{\hbar\omega_n}{2C_\mathrm{g}N}}\left(\hat a_{n}e^{i(k_nz-\omega_nt)}+\mathrm{H.c.}\right)
\end{equation}
where $\omega_n$ and $k_n$ are the angular frequency and wavenumber of the $n$-th mode while $\hat a_n$ is its annihilation  operator of the $n$-th mode. Positive indexes denote progressive waves ($k_n>0$ and $\omega_n>0$), while negative indexes denote regressive waves ($k_{-n}=-k_n<0$ and $\omega_{-n}=\omega_n$).\\
It is trivial to recover the link between the voltage drop and the current passing through a cell given by the classical Telegrapher's equation, which exploits the inductance of the cell for the $n$-th mode $L_{n}$
\begin{equation} \label{eq10}
    \frac{\partial V_n}{\partial z} = -\frac{L_n}{a}\frac{\partial I_n}{\partial t}
\end{equation}

$L_{n}$ can be calculated as the parallel between the effective inductance $L_{\mathrm{eff},n}$ (composed by the Josephson capacitance $C_\mathrm{J}$ and the geometrical inductance $L_\mathrm{g}$, see Supplemental \ref{appendix:effectiveinductance}) and the Josephson inductance $L_\mathrm{J}$. Exploiting the constitutive relation for a generic inductor it can be written that $\Delta{\Phi}=L\cdot I$. 
Hence, using the flux-current relation of a Josephson junction, $I_\mathrm{J}=I_\mathrm{c}\sin{\left(\Delta\Phi/\varphi_0\right)}$, the Josephson inductance $L_\mathrm{J}$ can be simply expressed, via a nonlinear relation with $\Delta\Phi$, as
\begin{equation}\label{eq:Lj}
L_\mathrm{J} =\frac{\Delta\Phi}{I_\mathrm{J}}= \frac{\varphi_0}{I_\mathrm{c}}\frac{\Delta\Phi/\varphi_0}{\sin{\left(\Delta\Phi/\varphi_0\right)}} \equiv L_{\mathrm{J}_0}\frac{\Delta\Phi/\varphi_0}{\sin{\left(\Delta\Phi/\varphi_0\right)}}
\end{equation}

with $L_{\mathrm{J}_0}=\varphi_0/I_\mathrm{c}$. It follows that the cell inductance $L_n$ can be written as
\begin{align}\label{eqLn}
L_n
&=\frac{\Lambda_{n}L_\mathrm{g}}{1+\Lambda_{n}\frac{L_\mathrm{g}}{L_{\mathrm{J}_0}}\frac{\sin{\left(\Delta\Phi/\varphi_{0}\right)}}{\left(\Delta\Phi/\varphi_{0}\right)}}
\end{align}

where the dispersion coefficient of the $n$-th node $\Lambda_n=1/(1-\omega_n^2L_\text{g}C_\text{J})$ (Supplemental \ref{appendix:effectiveinductance}) has been defined. The time-dependent component of equation \eqref{eq:2} can be found exploiting the mode decomposition for the AC current through the cell $I_n$ and the inductance $L_n$ for the corresponding mode as
\begin{equation}\label{eq4modes}
    \delta\Phi=\sum_n L_n I_n
\end{equation}

Replacing the classical variables by corresponding operator, and accordingly to the standard quantum description of electrical circuits \cite{Vool2017b}, it follows that (see Supplemental \ref{appendix:nonlinear_flux})
\begin{equation}\label{ACflux}
    \delta\hat{\Phi} =\sum_n\left[ \left(1+\Lambda_{n}\frac{L_\mathrm{g}}{L_{\mathrm{J}_0}}\frac{\sin{\left(\frac{\Delta\Phi_{\mathrm{DC}} + \delta\hat{\Phi}}{\varphi_{0}}\right)}}{\frac{\Delta\Phi_{\mathrm{DC}} + \delta\hat{\Phi}}{\varphi_{0}}}\right)^{-\frac{1}{2}}\delta\hat{\Phi}^{(0)}_{n}\right]
\end{equation}

where the zero order AC flux component of the $n$-th mode $\delta \hat \Phi_n^{(0)}$ has been defined. Equation \eqref{ACflux} is an implicit relation for the flux operator $\delta\hat \Phi$, which can be solved at zero order by the substitution $\delta\hat \Phi \mapsto \delta\hat \Phi^{(0)}$ in the right-hand side.\\
In order to find an analytical solution one can perform the Taylor expansion of the square root into equation \eqref{ACflux} and of the Josephson energy into \eqref{eq:first_quantization_hamiltonian} for $\delta\hat{\Phi}^{(0)}\ll \varphi_0$. The maximum order of expansion was chosen to take into account scattering events involving at most 4 photons. This procedure provides a valid approximation for the nonlinear time-dependent flux operator $\delta \hat \Phi$ that can be substituted into equation \eqref{eq:first_quantization_hamiltonian} to obtain the Hamiltonian of the system in terms of ladder operators.
\begin{align}\label{eq:H_Full_compact}
\hat{H}&=\hbar\chi_0+\sum_n\hbar\chi_1^{(n)}\left(\hat{a}^{\dag}_n\hat{a}_n+\frac{1}{2}\right)+ \nonumber \\
&+\sum_{n,l,m} \hbar\chi_3^{(n,l,m)}\left\{\hat{a}+\hat{a}^{\dag}\right\}_{n,l,m}\delta_{\Delta\omega_{n,l,m},\;0}+ \nonumber \\
&+\sum_{n,l,m,s} \hbar\chi_4^{(n,l,m,s)}\left\{\hat{a}+\hat{a}^{\dag}\right\}_{n,l,m,s}\delta_{\Delta\omega_{n,l,m,s},\;0}
\end{align}

The subscripts of the braces in equation \eqref{eq:H_Full_compact} stand for a multiplication of the form $\{\hat{a}+\hat{a}^\dagger\}_{n,l,...,k}=(\hat{a}_n+\hat{a}^\dagger_n)(\hat{a}_l+\hat{a}^\dagger_l)\cdot ... \cdot (\hat{a}_k+\hat{a}^\dagger_k) $. The $\delta_{\Delta \omega,0}$ Kronecker functions have the role to select the only scattering events that fulfill the energy conservation among the three ($\Delta\omega_{n,l,m}=0$) or four ($\Delta\omega_{n,l,m,s}=0$) modes taken into account (for an example see \footnote{$\Delta\omega_{n,l,m}=\pm\omega_n\pm\omega_l\pm\omega_m$, where the sign of each addend is defined by the combination of creation and annihilation operators that precedes this quantity (minus sign if related to a creation operator, plus sign if related to an annihilation operator). For instance $(\hat{a}_n^{\dag}e^{-i(k_nz-\omega_nt)})(\hat{a}_l^{\dag}e^{i(k_lz-\omega_lt)})(\hat{a}_m^{\dag}e^{-i(k_mz-\omega_mt)})=\hat{a}_n^{\dag}\hat{a}_l\hat{a}_m^{\dag}e^{i(\Delta k_{n,l,m}z-\Delta\omega_{n,l,m}t)}$, where $\Delta k_{n,l,m}=-k_n+k_l-k_m$ and $\Delta \omega_{n,l,m}=-\omega_n+\omega_l-\omega_m$}).\\
With this in hand, the full Hamiltonian of the system is found to be composed by a sum of four terms, the last two being interaction terms, where three or more modes give rise to 3WM or 4WM.\\
$\chi_1^{(n)}$ and $\chi_0$ describe respectively the free field energy of the traveling modes and the magnetic energy stored into the rf-SQUIDs due to the magnetic field or DC current bias applied. Furthermore, $\chi_3^{(n,l,m)}$ and $\chi_4^{(n,l,m,s)}$ are respectively the coupling parameters that characterize the 3WM and 4WM, both strongly dependent on the circuit parameters of the unit cell and on the frequency of the modes that populate the JTWPA. For the complete expression of the coupling coefficients as a function of the layout and experimental parameters see Supplemental \ref{app:CouplingCoefficients}. The distinctive characteristic of the layout under study is the strong dependence of these coupling parameters to the external bias conditions, opening the possibility to properly select a working regime (3WM or 4WM). Each coupling parameter, defined by a set of indices (e.g. $n$, $l$, $m$ and $s$), quantifies the interaction strength of the respective modes. It is then clear that different combinations of indices represent different effects that take place in the JTWPA which contribute to the output field. Focusing on a particular working regime of the amplifier it can be noted that if the JTWPA is biased so that the Kerr-like nonlinearity is suppressed, it is legitimate to consider the amplifier as a pure three-wave mixer \cite{Vissers2016}, where the conservation of energy imposes the creation of the so-called idler mode at frequency $\omega_\mathrm{p}-\omega$, being $\omega_\mathrm{p}$ and $\omega$ the pump and signal frequencies respectively.
On the contrary, if the quadratic nonlinearity gets suppressed, the JTWPA becomes a pure four-wave mixer, hence, with the degenerate pump assumption valid from now on for the 4WM regime (e.g. the two pump photons involved in the scattering have the same frequency), the idler will be located at $2\omega_\mathrm{p} - \omega$.

\begin{figure}[h]
    \centering
    \includegraphics[width=\linewidth]{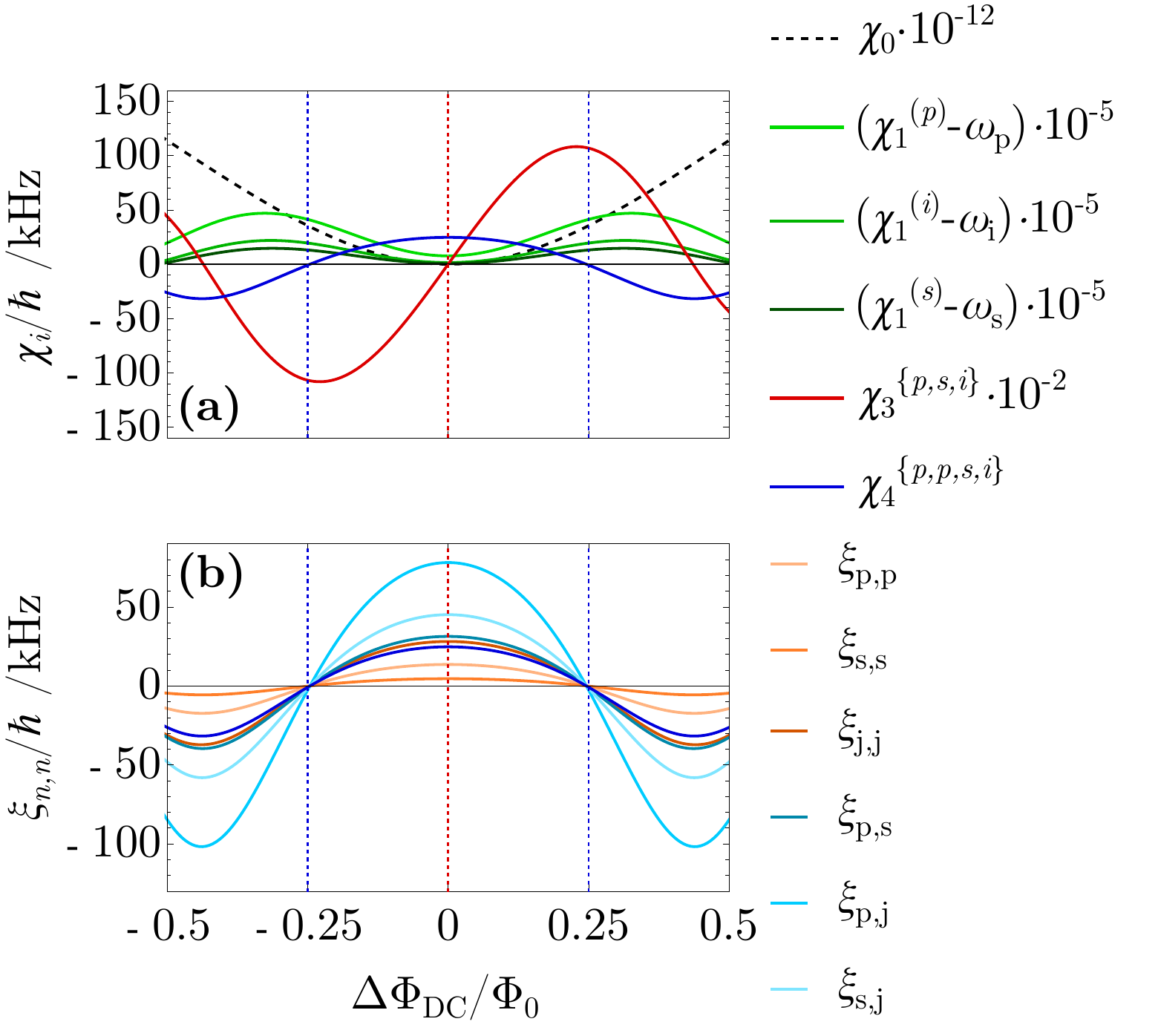}
    \caption{Hamiltonian coupling parameters $\chi_i$ characterizing the Hamiltonian \eqref{eq:H_Full_compact} versus the normalized DC flux bias ($\Delta\Phi_{\mathrm{DC}}/\Phi_0$). The Hamiltonian coefficient related to the constant flux bias ($\chi_0$) has been scaled by a factor of $10^{-12}$, the \textcolor{black}{non-interacting-modes Hamiltonian coupling constants ($\chi_1^{(n)}$)} have been shifted by the frequency of the corresponding photon ($\omega_\mathrm{i}$) and scaled by a factor $10^{-5}$ while $\chi_3^{(p,s,i)}$ has been scaled by a factor of $10^{-2}$. The indices in the superscripts vary with the considered mode and can take the values p (pump), s (signal) and i (idler). The blue vertical lines indicate the flux biases at which the amplifier works as a three-wave mixer, while the red vertical lines indicate the flux biases at which the amplifier works as a four-wave mixer (see Section \ref{sec:PDC} for a detailed description). The coupling parameters $\xi_{n,n}$ and $\xi_{n,l}$ refer to the self-phase and cross-phase modulation due to the 4WM interaction. The circuit parameters used to perform the numerical evaluations and plots are summarized in Table \ref{parameters_table}.
    }\label{fig:HamiltonianParametersPlot}
\end{figure}

\begin{table}[t]
\centering
\begin{tabular}{ |c|c|c|c| } 
\hline
Parameter & Value & Description\\
\hline
$I_\mathrm{c}$ & $\SI{5}{\micro\ampere}$ & Josephson critical current\\ 
$C_\mathrm{g}$ & $\SI{14}{\femto\farad}$ & Ground capacitance\\
$L_\mathrm{g}$ & $\SI{53}{\pico\henry}$ & Geometrical inductance\\
$C_\mathrm{J}$ & $\SI{60}{\femto\farad}$ & Josephson capacitance\\
$a$   & $\SI{60}{\micro\meter}$ & Unit cell length\\
$N$   & 900 & Number of unit cells\\
$\omega_\mathrm{p}$ & $2\pi\cdot \SI{12}{\giga\hertz}$ & Pump frequency\\
$\omega_\mathrm{s}$ & $2\pi\cdot \SI{7}{\giga\hertz}$ & Signal frequency\\
$\omega_\mathrm{i}$ & $2\pi\cdot \SI{5}{\giga\hertz}$ & 3WM idler frequency\\
$\omega_\mathrm{j}$ & $2\pi\cdot \SI{17}{\giga\hertz}$ & 4WM idler frequency\\
$\Delta\Phi_\mathrm{DC,3WM}/\Phi_0$ & $0.25$ & 3WM working point\\
$\Delta\Phi_\mathrm{DC,4WM}/\Phi_0$ & $0$ & 4WM working point\\
\hline
\end{tabular}
\caption{Circuit parameters and magnetic field flux bias (working points) used for numerical evaluations}
\label{parameters_table}
\end{table}

\subsection{\label{sec:PDC}Gain, Noise and Squeezing in 3WM and 4WM}

To analytically treat the problem the number of traveling modes that populate the JTWPA will be restricted to three, the input pump and signal frequencies plus the idler frequency that changes depending on the active nonlinearity. This assumption implies that mixed 3WM/4WM conditions will not be taken into account, these latter would require a four coupled modes discussion, that goes beyond the scope of this paper. Furthermore, since now the 3WM regime will be considered in a non-degenerate condition, that is $\omega\neq\omega_\mathrm{p}/2$. This regime is insensitive to the phase difference between the incoming waves \cite{vanderreep}, making the non-degenerate parametric amplifier $\textit{phase-preserving}$. With this in hand the full Hamiltonian \eqref{eq:H_Full_compact}, can be reduced to two different forms depending on the regime the amplifier is working in. Concerning the 3WM, the following Hamiltonian is obtained 
\begin{align}\label{eq:H3WM_Full}
    \hat{H}_{\mathrm{3WM}}&=\hbar \chi_0 +
    \sum_{\substack{n= \\ \{\omega_\mathrm{p},\omega,\omega_\mathrm{p}-\omega\}}}\hbar \chi_1^{(n)}\left(\hat{a}^{\dagger}_n\hat{a}_n+\frac{1}{2}\right)+\nonumber\\
    &+\hbar\chi_3^{\{\omega_\mathrm{p},\omega,\omega_\mathrm{p}-\omega\}}\left(\hat{a}^{\dagger}_{\omega_\mathrm{p}}\hat{a}_{\omega}\hat{a}_{\omega_\mathrm{p}-\omega}+\hat{a}^{\dagger}_{\omega}\hat{a}^{\dagger}_{\omega_\mathrm{p}-\omega}\hat{a}_{\omega_\mathrm{p}}\right)
\end{align}

having introduced $\chi_3^{\{\omega_\mathrm{p},\omega,\omega_\mathrm{p}-\omega\}}$ as the sum of all the possible terms arising from index permutations of $\chi_3^{(\omega_\mathrm{p},\omega,\omega_\mathrm{p}-\omega)}$ neglecting permutations signs degeneracy. While the 4WM Hamiltonian results to be
\begin{align}\label{eq:H4WM_Full}
    \hat{H}_{\mathrm{4WM}}&=\hbar \chi_0 +\hbar\xi_{0}+
    \sum_{\substack{n= \\ \{\omega_\mathrm{p},\omega,2\omega_\mathrm{p}-\omega\}}}\hbar \chi_1^{(n)}\left(\hat{a}^{\dagger}_n\hat{a}_n+\frac{1}{2}\right)+\nonumber\\
    &+\sum_{\substack{n= \\ \{\omega_\mathrm{p},\omega,2\omega_\mathrm{p}-\omega\}}}\hbar \xi_n \hat a_n^\dagger \hat a_n +\nonumber\\
    &+\sum_{\substack{n,l= \\ \{\omega_\mathrm{p},\omega,2\omega_\mathrm{p}-\omega\}}}\hbar \xi_{n,l}\hat a_n^\dagger \hat a_n \hat a_l^\dagger \hat a_l+\nonumber\\[1em]
    &+ \hbar \chi_4^{\{ \omega_\mathrm{p},\omega_\mathrm{p},\omega,2\omega_\mathrm{p}-\omega \}}\cdot\nonumber\\
    &\cdot\left(\hat{a}_{\omega_\mathrm{p}}^\dagger \hat a_{\omega_\mathrm{p}}^\dagger \hat a_{\omega} \hat a_{2\omega_\mathrm{p}-\omega} + \hat{a}_{\omega}^\dagger \hat a_{2\omega_\mathrm{p}-\omega}^\dagger \hat a_{\omega_\mathrm{p}} \hat a_{\omega_\mathrm{p}}\right)
\end{align}

where $\xi_0$ is a small correction to the zero-point energy, $\xi_n$ is a small contribution to the free-field energy of the modes and $\xi_{n,l}$ is the coefficient describing the self- ($n=l$) and cross-phase ($n \neq l$) modulation phenomena. Likewise the 3WM case, $\chi_4^{\{\omega_\mathrm{p},\omega_\mathrm{p},\omega,2\omega_\mathrm{p}-\omega\}}$ is the sum of all the possible terms that derive from index permutations of $\chi_4^{(\omega_\mathrm{p},\omega_\mathrm{p},\omega,2\omega_\mathrm{p}-\omega)}$ neglecting permutations signs degeneracy.\\
Figure \ref{fig:HamiltonianParametersPlot} shows the behaviour of the most significant coupling parameters as a function of $\Delta \Phi_{\mathrm{DC}}$. These coefficients present a periodic behaviour given by the periodicity of the Josephson inductance and the red and blue vertical lines represent particular bias values (working points) that select the 4WM or 3WM working regimes respectively. From now on we recall the 3WM and 4WM regimes by referring respectively to the blue and red vertical lines in the positive $\Delta \Phi_\mathrm{DC}$ plane of Figure \ref{fig:HamiltonianParametersPlot}. For their numerical values see Table \ref{parameters_table}.\\
Once the Hamiltonian of the system is known, it is possible to determine the dynamic of the observables. Exploiting the Heisenberg picture of quantum mechanics, the time evolution of the creation and annihilation operators can be computed through the Heisenberg equation $\mathrm{d}\hat{a}_\mathrm{H}(t)/\mathrm{d}t = (i/\hbar) [\hat{H},\hat{a}_\mathrm{H}(t)] + (\partial \hat{a}/\partial t)_{\mathrm{H}}$ (for the complete calculations see Supplemental \ref{appendix:coupled_mode_equations}). From here to the end of Section \ref{sec:PDC} we will drop the $\mathrm{H}$ subscript.\\
From the calculation of the Heisenberg equations a system of coupled equations for the creation and annihilation operators (3WM \eqref{eq:dap_on_dt_3WM}-\eqref{eq:dai_on_dt_3WM}, 4WM \eqref{eq:dap_on_dt_4WM}-\eqref{eq:dai_on_dt_4WM}) comes out. This system is in general not solvable analytically unless one performs some approximations \cite{Gambini1977}. Indeed, one can proceed with the so-called \emph{undepleted pump approximation} to analytically treat the system. Such an approximation requests the pump amplitude to be much higher than the signal and idler ones so that its magnitude does not change significantly during the interaction process. On the other hand, under the so-called \emph{classical pump approximation}, the ladder operator describing the pump mode can be treated as a classical amplitude 
\begin{align}\label{eq:Ap}
    \sqrt{\frac{2\hbar\omega_\mathrm{p}}{C_\mathrm{g}N}}\hat{a}_\mathrm{p} \mapsto A_\mathrm{p}
\end{align}

being $A_\mathrm{p}$ the classical voltage amplitude of $\hat{V}_{C_\mathrm{g}}$ (equation \eqref{eq9}).\\
The strong interplay between the traveling waves manifests itself in a system of coupled differential equations for the annihilation operators describing the signal and idler modes
\begin{subequations}
\begin{eqnarray}
 \frac{\mathrm{d}\hat{a}_\omega}{\mathrm{d}t}
    &=-i\Upsilon \hat a_{\omega'}^\dagger e^{-i\Psi t} \label{eq:CME_signal}
\\
\frac{\mathrm{d}\hat{a}_{\omega'}}{\mathrm{d}t}
    &=-i\Upsilon \hat a_{\omega}^\dagger e^{-i\Psi t} \label{eq:CME_idler}
\end{eqnarray}
\end{subequations}
where the \emph{density phase mismatch} $\Psi$ has been defined (3WM - equation \eqref{eq:phase_mismatch_density_3WM}, 4WM - equation \eqref{eq:phase_mismatch_density_4WM}). In equation \eqref{eq:CME_idler} the subscript $\omega'$ stands for a generic idler tone, both for the 3WM and the 4WM case. The main structure of the system remains the same regardless of the kind of interaction that takes place into the JTWPA, indeed it is possible to define an \emph{interaction parameter} $\Upsilon=\Upsilon_{\mathrm{3WM}, \mathrm{4WM}}$ that characterizes the working regime the amplifier is biased in
\begin{subequations}
\begin{eqnarray}
    \Upsilon_{\mathrm{3WM}} &= \chi_3|A_{\mathrm{p},0}|\\
    \Upsilon_{\mathrm{4WM}} &= \chi_4|A_{\mathrm{p},0}|^2
\end{eqnarray}
\end{subequations}

$\chi_{3,4}$ are two bias tunable coefficients that incorporate information about the strength of the quadratic or cubic non-linearity into the device (for their definition refer to equations \eqref{eq_chi4} and \eqref{eq_chi3}). It has to be noticed that in $\Upsilon_{\mathrm{3WM},\mathrm{4WM}}$ the proportionality to the initial pump amplitude $A_{\mathrm{p},0}$ reflects the nature of the scattering taken into account, hence involving one (linear) or two (quadratic) pump photons.\\
Under the undepleted pump assumption and working in the co-rotating frame one can find the following analytical solution to equations \eqref{eq:CME_signal} and \eqref{eq:CME_idler}
\begin{align}\label{eq49}
    \hat{a}_{\omega}(t)&=\left[ \hat{a}_{\omega,0}\Bigg(\cosh{(gt)}+\frac{i\Psi}{2g}\sinh{(gt)}\right)-\nonumber\\
    &\hspace{0.5cm}-\frac{i\Upsilon}{g}\left(\hat{a}_{\omega',0}\right)^\dagger\sinh{(gt)}\Bigg]e^{-i\left(\Psi/2\right)t}
\end{align}

being $\hat{a}_{\omega,0}$ and $(\hat{a}_{\omega',0})^\dagger$ the ladder operators at the initial interaction time and with the \textit{complex gain factor}
\begin{align}\label{eq:complexGain}
    g = \sqrt{\Upsilon^2-\left(\frac{\Psi}{2}\right)^2}
\end{align}

For the 3WM case, under experimentally reasonable parameters a \textit{negligible total phase mismatch approximation}, hence the phase mismatch density times the interaction time, can be considered in equation \eqref{eq49}, so that $\Psi t\approx0$ and the phase lag between the traveling modes can be neglected. 
Moreover, under the undepleted pump approximation, it can be shown that the gain variation given by the phase mismatch density in \eqref{eq:complexGain} can be neglected since $\Upsilon^2 \gg \frac{\Psi^2}{4}$, giving the much simpler relation
\begin{align}\label{eq:complexGainNoMismatch_3WM}
    g &\approx |\Upsilon_{\mathrm{3WM}}|
\end{align}

It is helpful to introduce a set of auxiliary functions that incorporates the behaviour of the JTWPA and simplifies the notation

\begin{align}
 u(\omega,t)&=\cosh{(g(\omega)t)}+\frac{i\Psi(\omega)}{2g(\omega)}\sinh{(g(\omega)t)}
 \\
 v(\omega,t)&=-\frac{\Upsilon}{g(\omega)}\sinh{(g(\omega)t)}
\end{align}

By making use of \eqref{eq49} it is now possible to define the number of output signal photons as the average number of photons of frequency $\omega$ after a certain amount of time $t$ spent into the medium
\begin{widetext}
\begin{align}\label{eq:Gain_signal}
    \braket{\hat n_\mathrm{\omega}}=\braket{\hat{a}_\omega^{\dag}\hat{a}_\omega}=|u|^2\braket{\left(\hat{a}_{\omega,0}\right)^{\dag}\hat{a}_{\omega,0}}+
    |v|^2\left[
    \langle(\hat{a}_{\omega',0})^\dagger\hat{a}_{\omega',0}\rangle+1\right]+iu^*v
    \langle(\hat{a}_{\omega,0})^\dagger(\hat{a}_{\omega',0})^\dagger\rangle-iuv^*\langle\hat{a}_{\omega',0}\hat{a}_{\omega,0}\rangle \nonumber\\
\end{align}
\end{widetext}

Equation \eqref{eq:Gain_signal} is a general relation to estimating the number of outgoing signal photons regardless of the nature of the incoming state (Fock, coherent, thermal, etc.).\\
A parametric amplifier is a particular realization of a linear amplifier, of which the typical output field can be expressed as $\hat a_\mathrm{\omega} = \sqrt{G}\hat a_\mathrm{\omega, 0}+\hat L^{\dagger}$ (Equation ($2.9$) in \cite{caves2012}), hence as the sum of the input field times a real constant plus an additional operator. $\sqrt{G}=u$ is called the \textit{amplitude gain} of which the linear amplifier increment of the input signal, while $\hat L=-iv^* \hat{a}_{\omega',0}e^{i(\Psi/2) t}$ is the \textit{added noise operator}, that is a property of the sole internal degrees of freedom of the amplifier. It is then straightforward to rewrite \eqref{eq:Gain_signal} as
\begin{align} \label{eq:linear_ampl_output}
    \braket{\hat n_\mathrm{\omega}} &= G\braket{\left(\hat{a}_{\omega,0}\right)^{\dag}\hat{a}_{\omega,0}}+\nonumber\\ 
    & +\braket{\hat L^{\dagger}\hat L}+
    \sqrt{G}\left(\braket{\left(\hat{a}_{\omega,0}\right)^{\dag}\hat{L}^\dagger}+
    \braket{\hat{L}\hat{a}_{\omega,0}}\right)=\nonumber\\
    & =G\braket{\hat n_\mathrm{\omega,0}}+\braket{\hat{\mathcal{N}}}
\end{align}

where $\hat n_\mathrm{\omega,0}$ is the input signal photon number operator. Here one can recognize two key features of a linear amplifier: the \textit{photon number gain} $G$, that is the contribution to the total number of output photons given by the sole input field, and the \textit{noise photon number} operator $\braket{\hat{\mathcal{N}}}$, which embeds the contribution given by the amplifier itself. If the input idler mode is in the vacuum state the system acts as an ideal linear amplifier, hence characterized by a quantum-limited added noise \cite{caves2012}. By comparing Equations \eqref{eq:Gain_signal} and \eqref{eq:linear_ampl_output} one can write the gain and the noise photon number of a JTWPA as
\begin{align}
    G& =|u|^2=\cosh{gt}^2+\frac{\Psi^2}{4g^2}\sinh{gt}^2+\frac{i\Psi}{2g}\sinh{gt}\cosh{gt}\label{eq:gain}\\
    \braket{\hat{\mathcal{N}}}&=|v|^2\left[
    \langle(\hat{a}_{\omega',0})^\dagger\hat{a}_{\omega',0}\rangle+1\right]+\nonumber\\ 
    & +\:iu^*v
    \langle(\hat{a}_{\omega,0})^\dagger(\hat{a}_{\omega',0})^\dagger\rangle-iuv^*\langle\hat{a}_{\omega',0}\hat{a}_{\omega,0}\rangle \label{eq:added_noise}
\end{align}

and using the above relations the \textit{added-noise number} of the amplifier can be defined as the ratio between the symmetric variance of the added noise $\hat L$ and the gain $G$ \cite{caves2012} 

\begin{subequations}
\begin{eqnarray}
   \mathcal{A}&=&\frac{\braket{| \Delta \hat L |^2}}{G}\\
    &=&\frac{\braket{|\hat L|^2}-|\braket{\hat L}|^2}{G}\\
    &=&\frac{|v|^2}{|u|^2}\left( \frac{1}{2}+ \braket{\hat a^\dagger_\mathrm{\omega',0}a_\mathrm{\omega',0}}-|\braket{\hat a_\mathrm{\omega',0}}|^2\right)
\end{eqnarray}
\end{subequations}

In the case where the phase mismatch can be neglected $\Upsilon^2 \gg \frac{\Psi^2}{4}$ (this happens for low values of $C_\mathrm{J}$ and $L_\mathrm{g}$, which reduce the chromatic dispersion in the line) the gain becomes
\begin{align}\label{eq:Gain_signalNoidler}
    G& \approx \cosh^2{gt}
\end{align}

in accordance with the classical approach given by \cite{Zorin2016}.\\
It is now worth making few observations on equations \eqref{eq:gain} and \eqref{eq:added_noise}. The gain $G$ only depends on the layout of the amplifier and on the amplitude and frequency of the pump tone, as expected from a linear amplifier; in other words the gain does not depend on the input state. The expected value of the noise photon number operator ($\braket{\hat{\mathcal{N}}}$) has a non-trivial dependence on the annihilation and creation operators, and it is interesting evaluating this quantity for two simple cases: a Fock state $\ket{\psi_\mathrm{F}}=\ket{N^{\mathrm{S}}_{\mathrm{in}}}_\mathrm{s}\ket{N^{\mathrm{I}}_{\mathrm{in}}}_\mathrm{i}$ and a coherent  state $\ket{\psi_\mathrm{c}}=\ket{\alpha}_\mathrm{s}\ket{\beta}_\mathrm{i}$
\begin{align}
    \braket{\hat{\mathcal{N}}}_{\mathrm{F}}&=|v|^2\left(1+N^{\mathrm{I}}_{\mathrm{in}}\right)\\
    \braket{\hat{\mathcal{N}}}_{\mathrm{C}}&=|v|^2\left(1+|\beta|^2\right)-iuv^* \alpha\beta + i u^* v \alpha ^* \beta ^*
\end{align}

\begin{figure}[h] 
    \centering
    \includegraphics[width=0.9\linewidth]{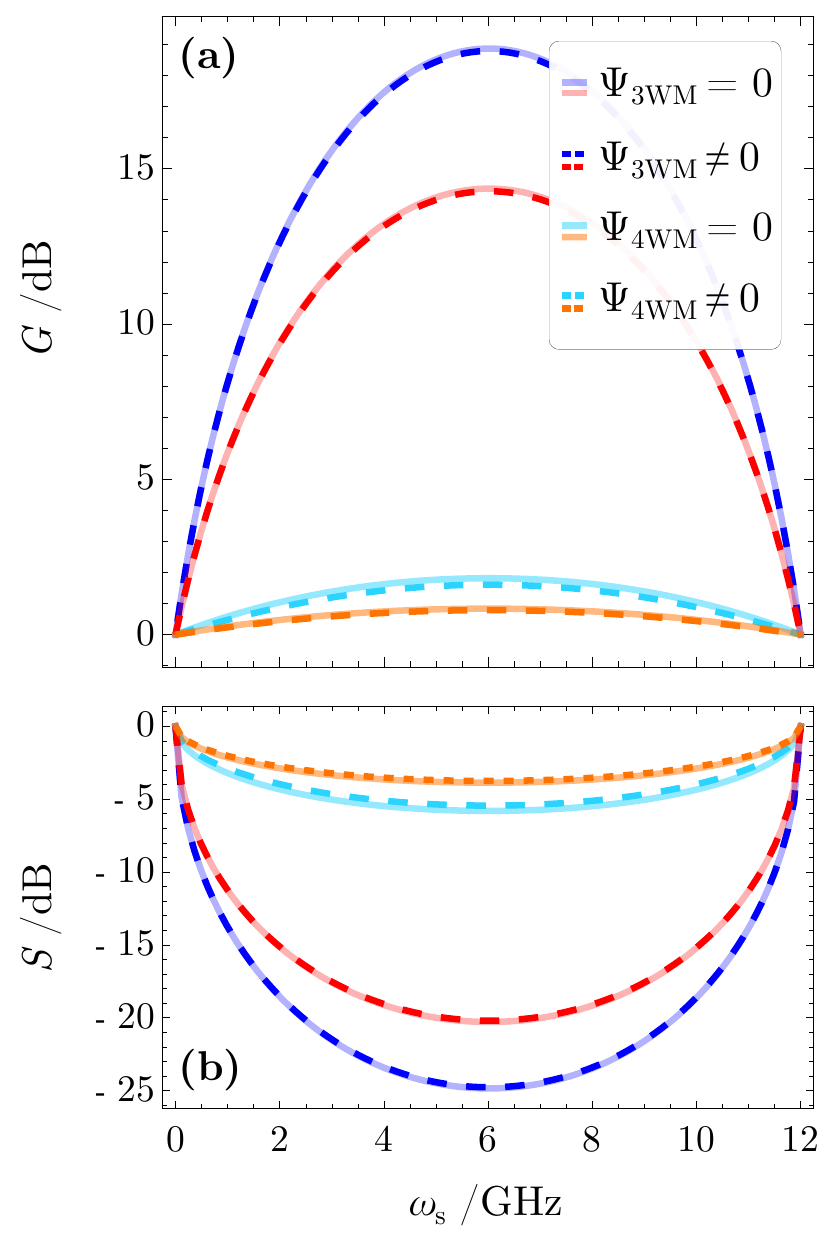}
    \caption{(a) Photon number gain $G$ of the JTWPA under the undepleted pump approximation expressed by equation \eqref{eq:gain} in the 4WM (light blue/orange) and 3WM (blue/red) regimes. For each pump power one can see the cases with and without zero phase mismatch, respectively $\Psi = 0$ and $\Psi \neq 0$. 
    (b) Squeezing spectrum $S$ (equation \eqref{eq:S_SqueezSpect_Vac_Psi=0}) as a function of the signal frequency calculated for a vacuum input state in the 4WM and 3WM regimes. Different colors express different pump currents ($I_\mathrm{p}$) for which $G$ and $S$ are calculated: blue/light blue $I_\mathrm{p}/I_\mathrm{c}=0.99$ ($P_\mathrm{p}=\SI{-64.6}{\decibel m}$), red/orange $I_\mathrm{p}/I_\mathrm{c}=0.81$ ($P_\mathrm{p}=\SI{-66.3}{\decibel m}$). The working points in the 3WM and 4WM cases are respectively $\Delta\Phi_\mathrm{DC,3WM}/\Phi_0=0.25$ and  $\Delta\Phi_\mathrm{DC,4WM}/\Phi_0=0$. Refer to Table \ref{parameters_table} for the experimental parameters used in the computations.}
    \label{fig:GSPlotFreq}
\end{figure}

Regardless of its non-trivial dependence, it turns out that if the input idler mode is in its vacuum state ($N^{\mathrm{I}}_{\mathrm{in}}=\beta=0$) the noise photon number simplifies becoming just $\braket{\hat{\mathcal{N}}}_{\mathrm{F}}=\braket{\hat{\mathcal{N}}}_{\mathrm{C}}=|v|^2$.\\
Figure \ref{fig:GSPlotFreq} (a) shows the photon number gain as a function of the signal frequency in 3WM and 4WM regimes at different pump powers. The curves, representing $G$ with and without negligible phase mismatch, show that the approximation $\Psi \approx 0$ holds in all the bandwidth for the 3WM and the 4WM regimes, for the set of parameters reported in Table \ref{parameters_table}.

The correlation of the signal and idler photons results in a squeezed output field of the JTWPA. To model these correlations, one can introduce quadratures as
\begin{align}\label{eq:Y_Quadratures}
    \hat{Y}^{\theta}(\omega)=i(e^{i\theta /2}\hat{a}_{\omega}^{\dagger} - e^{-i\theta /2}\hat{a}_{\omega} )
\end{align}

with their associated fluctuations
\begin{align}\label{quadrature_fluctuation}
    \Delta\hat{Y}^{\theta}(\omega)=\hat{Y}^{\theta}(\omega)-\braket{\hat{Y}^{\theta}(\omega)}
\end{align}
being $\theta$ the so-called squeezing angle. From the previous definitions, one can compute (see Supplemental \ref{appendix:squeezing}) the relation between the squeezing spectrum $S$ and the quadratures fluctuations as
\begin{align}\label{eq:S_SqueezingSpectrum}
    S(\omega)=\sum_n \braket{\Delta\hat{Y}^{\theta}(\omega) \Delta\hat{Y}^{\theta}(\omega_n)}
\end{align}

For a vacuum input state, it can be shown that the product of the fluctuations of the two quadratures gives the minimum possible value allowed by the Heisenberg uncertainty principle, the fingerprint of a quantum-limited amplification \cite{PhysRevLett.112.190504}. From \eqref{eq:S_SqueezingSpectrum} the squeezing spectrum is then
\begin{align}\label{eq:S_SqueezSpect_Vac_Psi=0}
    S(\omega)&= 1 + 2|v(\omega, t)|^2 - 2|v(\omega, t)|\sqrt{|v(\omega, t)|^2 + 1}
\end{align}

\begin{figure}[h] 
    \centering
    \includegraphics[width=8cm]{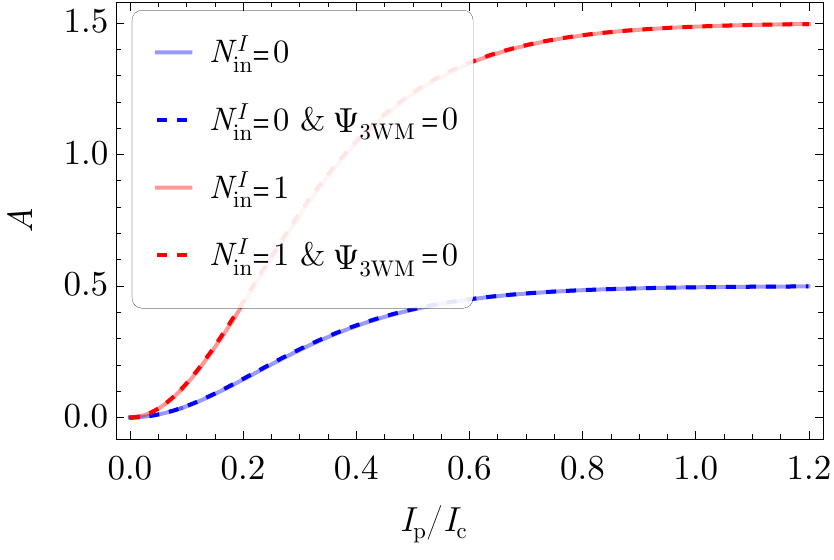}
    \caption{Added-noise number ($\mathcal{A}$) as a function of the normalized pump current, for a pump frequency of $\SI{12}{GHz}$ and a signal frequency of $\SI{7}{GHz}$ in 3WM working point. The dashed curves are calculated in case of negligible phase mismatch while the solid curves in the case of non-negligible phase mismatch. The blue curves are considered for an input Fock state where the idler is in the vacuum state ($\ket{N^{\mathrm{S}}_{\mathrm{in}},0}$), while the red curves for an input Fock state with one idler photon ($\ket{N^{\mathrm{S}}_{\mathrm{in}},1}$).}
    \label{fig:noise_figure}
\end{figure}

Figure \ref{fig:GSPlotFreq}(b) shows the squeezing spectrum of equation \eqref{eq:S_SqueezSpect_Vac_Psi=0} plotted as a function of the signal frequency for different input pump powers, calculated for a vacuum input state.\\
The behaviour of the amplifier is determined not only by its constructive parameters but also from some dynamical features like the pump and signal power and frequency that can by adjusted during the experiment. An example of this can be appreciated by evaluating the added-noise number $\mathcal{A}$ as a function of the pump power.
Figure \ref{fig:noise_figure} reports $\mathcal{A}$ in the 3WM working point, for a Fock input state in the case of one and no idler input photons (respectively $\ket{N^{\mathrm{S}}_{\mathrm{in}},1}$ and $\ket{N^{\mathrm{S}}_{\mathrm{in}},0}$) both for negligible and non-negligible phase mismatch as a function of the pump current. For an initial idler vacuum state ($\ket{N^{\mathrm{S}}_{\mathrm{in}},0}$) the value of $\mathcal{A}$ grows from zero and saturates up to the value $0.5$ for high pump currents both for negligible and non-negligible phase mismatch, as expected from a quantum limited amplifier. It is straightforward considering an input state with one idler photon ($\ket{N^{\mathrm{S}}_{\mathrm{in}},1}$). In that case the added-noise number saturates at $1.5$ for high pump currents, hence the value coming from standard quantum limit ($0.5$) plus the number of incoming idler photons ($1$).

\subsection{Interaction of quantum states through 3WM or 4WM}\label{interaction_of_quantum_states}

The time evolution of the state vectors can give important hints on the behaviour of the JTWPA in presence of single-photon signals. Moving to the framework of the interaction picture it is possible to calculate the output photon statistics in the Fock base for any incoming state. The time evolution of a quantum state $\ket{\psi(t)}$ can be expressed as
\begin{equation}
\label{eq:time_evolution_quantumstate_1}
    \ket{\psi(t)}=e^{-\frac{i}{\hbar}\int_0^t\hat{H}_{\text{int}}\mathrm{d}t'}\ket{\psi(0)}
\end{equation}

where $\hat{H}_{\text{int}}=\hat{H}_{\text{int,3WM(4WM)}}$ is the Three-, Four-Wave Mixing Hamiltonian written in the co-rotating frame under the undepleted pump approximation:
\begin{align}
\hat{H}_{\text{int,3WM}}=\hbar\chi_3|A_{\mathrm{p},0}|\left(\hat{a}_\mathrm{s}\hat{a}_\mathrm{i}e^{i\Psi t}+\hat{a}_\mathrm{s}^{\dag}\hat{a}_\mathrm{i}^{\dag}e^{-i\Psi t}\right) \label{appa}
\\
\hat{H}_{\text{int,4WM}}=\hbar\chi_4|A_{\mathrm{p},0}|^2\left(\hat{a}_\mathrm{s}\hat{a}_\mathrm{i}e^{i\Psi t}+\hat{a}_\mathrm{s}^{\dag}\hat{a}_\mathrm{i}^{\dag}e^{-i\Psi t}\right) \label{appb}
\end{align}
\\

To give an analytical solution of the problem we consider a set of parameters where the negligible phase mismatch condition can be considered valid (i.e., $\Psi t\ll 1$). Under this assumption equation \eqref{eq:time_evolution_quantumstate_1} becomes
\begin{equation}
    \label{eq:time_evolution_quantumstate_2}
    \ket{\psi(t)}=e^{i\kappa\left(\hat{a}_\mathrm{s}\hat{a}_\mathrm{i}+\hat{a}_\mathrm{s}^{\dag}\hat{a}_\mathrm{i}^{\dag}\right)}\ket{\psi(0)}
\end{equation}

where $\kappa=-\chi_3|A_{\mathrm{p},0}|t$ ($\kappa=-\chi_4|A_{\mathrm{p},0}|^2t$) is the amplification factor for the 3WM (4WM) case. Equation \eqref{eq:time_evolution_quantumstate_2} can be written in a normal ordered form \cite{Barnett1997} as 
\begin{align}
    \label{eq:time_evolution_quantumstate_3}
    \ket{\psi(t)}=&e^{i\tanh{(\kappa)}\hat{a}_\mathrm{s}^{\dag}\hat{a}_\mathrm{i}^{\dag}}\cdot\nonumber\\
    &\cdot e^{-\ln{[\cosh{(\kappa)}]\left(1+\hat{a}_\mathrm{s}^{\dag}\hat{a}_\mathrm{s}+\hat{a}_\mathrm{i}^{\dag}\hat{a}_\mathrm{i}\right)}}\cdot\nonumber\\
    &\cdot e^{i\tanh{(\kappa)}\hat{a}_\mathrm{s}\hat{a}_\mathrm{i}}\ket{\psi(0)}
\end{align}

In the following, the time evolution of two different classes of initial input states will be analyzed. 

\subsubsection{Fock States input}

This subsection focuses on the time-evolution of an initial Fock state $\ket{\psi_\mathrm{F}(0)}=\ket{N^{\mathrm{S}}_{\mathrm{in}}}_\mathrm{s}\ket{N^{\mathrm{I}}_{\mathrm{in}}}_\mathrm{i}$. Considering the action of Equation \eqref{eq:time_evolution_quantumstate_3} on the initial state, by means of a power expansion of each exponential function, the expression of the quantum state at a certain time $t$ can be derived.\\
Then, the expectation value of the signal photon number operator $\hat{n}_\mathrm{s}=\hat{a}_\mathrm{s}^{\dag}\hat{a}_\mathrm{s}$ on the final state $\ket{\psi_\mathrm{F}(t)}$ can be expressed as 
\begin{equation}
\label{eq:ns_expect_Fock}
    \braket{\hat{n}_\mathrm{s}}_{\psi_\mathrm{F}(t)}=\braket{\psi_\mathrm{F}(t)|\hat{n}_\mathrm{s}|\psi_\mathrm{F}(t)}=\sum_{N^{\mathrm{S}}_{\mathrm{fin}}}P_\mathrm{F}\left(N^{\mathrm{S}}_{\mathrm{fin}}\right)\cdot N^{\mathrm{S}}_{\mathrm{fin}}
\end{equation}

where $P_\mathrm{F}(N^{\mathrm{S}}_{\mathrm{fin}})$ is the probability to measure $N^{\mathrm{S}}_{\mathrm{fin}}$ signal photons in the final state, and $N^{\mathrm{S}}_{\mathrm{in}}-\min{\{N^{\mathrm{S}}_{\mathrm{in}},N^{\mathrm{I}}_{\mathrm{in}}\}}<N^{\mathrm{S}}_{\mathrm{fin}}<\infty$. This normalized probability distribution can be expressed by exploiting the binomial coefficients as a function both of the characteristics of the initial state and of the characteristics of the medium
\begin{align}
    \label{eq:Probability_Distribution_Fock}
    P_\mathrm{F}&=\sum^{\min{\{N^{\mathrm{S}}_{\mathrm{in}},N^{\mathrm{I}}_{\mathrm{in}}\}}}_{n,n'=0}\frac{(-1)^{n-n'}\left[\tanh{(\kappa)}\right]^{2(N^{\mathrm{S}}_{\mathrm{fin}}-N^{\mathrm{S}}_{\mathrm{in}}+n+n')}}{\left[\cosh{(\kappa)}\right]^{2(1+N^{\mathrm{S}}_{\mathrm{in}}+N^{\mathrm{I}}_{\mathrm{in}}-n-n')}}\cdot\nonumber\\
    &\hspace{0.75cm}\cdot \binom{N^{\mathrm{S}}_{\mathrm{in}}}{n'}\binom{N^{\mathrm{I}}_{\mathrm{in}}}{n}\binom{N^{\mathrm{S}}_{\mathrm{fin}}}{N^{\mathrm{S}}_{\mathrm{in}}-n}\binom{N^{\mathrm{S}}_{\mathrm{fin}}-N^{\mathrm{S}}_{\mathrm{in}}+N^{\mathrm{I}}_{\mathrm{in}}}{N^{\mathrm{I}}_{\mathrm{in}}-n'}
\end{align}

\begin{figure}
    \centering
    \includegraphics[width=\linewidth]{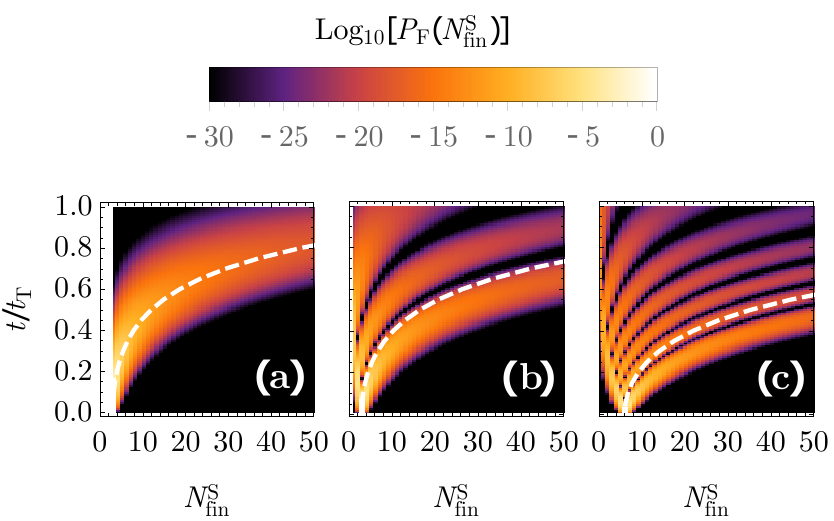}
    \caption{Time evolution inside the medium, from its input port ($t=0$) to the output port ($t=t_T$) of the probability distribution $P_\mathrm{F}$ to find $N^{\mathrm{S}}_{\mathrm{fin}}$ signal photons for three different initial Fock states (a)$\ket{3}_\mathrm{s}\ket{0}_\mathrm{i}$, (b)$\ket{3}_\mathrm{s}\ket{2}_\mathrm{i}$ and (c) $\ket{6}_\mathrm{s}\ket{6}_\mathrm{i}$ calculated for $I_\mathrm{p}=0.5I_\mathrm{c}$. The dashed white lines represent the time evolution of the expectation value $\braket{\hat{n}_\mathrm{s}}$.}
    \label{fig:PFockVsTimeXNSinXNIinPlot}
\end{figure}

In figure (\ref{fig:PFockVsTimeXNSinXNIinPlot}) the time evolution of the probability distribution is represented for three different initial number states, for a 3WM interaction and for the experimental parameters reported in Table \ref{parameters_table}. In all cases, at the beginning of the interaction, the probability distribution is single-peaked and has a maximum in correspondence of $N^{\mathrm{S}}_{\mathrm{in}}$, then the distribution can turn into a multi-peaked distribution if $N^{\mathrm{I}}_{\mathrm{in}}\neq0$. In this case, the distance between peaks increases with time. After the initial transition time, the number of maxima becomes constant and equal to  $\min\{N^{\mathrm{S}}_{\mathrm{in}},N^{\mathrm{I}}_{\mathrm{in}}\}+1$. This value reflects the number of possible combinations that can occur between the initial signal and idler photons at the beginning of the interaction. Considering the case of an initial state $\ket{3}_\mathrm{s}\ket{2}_\mathrm{i}$ (Fig.\;\ref{fig:PFockVsTimeXNSinXNIinPlot}(b)), at $t=0$ and with a certain probability, two couples of signal-idler photons may virtually recombine to create a pair of pump photons. This leads to the \textit{effective} propagation and amplification of a single remaining signal photon. Yet, with a different given probability, just a single couple of signal-idler photons or none of them virtually recombine, leading to the \textit{effective} propagation and amplification of, respectively, two or three signal photons. In addition, considering Fig.\;\ref{fig:PFockVsTimeXNSinXNIinPlot}(c) it can be noted that in the case of $N^{\mathrm{S}}_{\mathrm{in}}=N^{\mathrm{I}}_{\mathrm{in}}$, so an equal number of signal and idler photons, despite the the fact that $N^{\mathrm{S}}_{\mathrm{in}}\neq0$, the probability to observe at the end of the amplifier a vacuum state is significantly non-zero. This is in accordance with the fact that an \textit{effective} propagation and amplification of the vacuum state can occur.\\
\subsubsection{Coherent States input}
Similarly to what has been performed in the case of a Fock state input, the expectation value of the signal photon number operator can be derived by \eqref{eq:ns_expect_Fock} considering an initial bimodal coherent state $\ket{\psi_\mathrm{c}(0)}=\ket{\alpha}_\mathrm{s}\ket{\beta}_\mathrm{i}$. The calculation leads to the following probability distribution 
\begin{align}
    P_\mathrm{C}&=\sum^{\infty}_{m,n,n'=0}\frac{(-1)^{n-n'}\left[\tanh{(\kappa)}\right]^{n+n'}}{\left[\cosh{(\kappa)}\right]^{2(1+N^{\mathrm{S}}_{\mathrm{fin}}+m-n')}}\cdot\nonumber\\
    &\hspace{0.75cm}\cdot\frac{\alpha^{N^{\mathrm{S}}_{\mathrm{fin}}-n}\left(\alpha^{*}\right)^{N^{\mathrm{S}}_{\mathrm{fin}}-n'}\beta^m\left(\beta^{*}\right)^{m+n-n'}}{e^{\left[|\alpha|^2+|\beta|^2+i(\alpha^{*}\beta^{*}-\alpha\beta)\tanh{(\kappa)}\right]}}\cdot\nonumber\\
    &\hspace{0.75cm}\cdot\frac{1}{m!\;(N^{\mathrm{S}}_{\mathrm{fin}}-n)!}\binom{N^{\mathrm{S}}_{\mathrm{fin}}}{n}\binom{m+n}{n'}
\end{align}

\begin{figure}
    \centering
    \includegraphics[width=\linewidth]{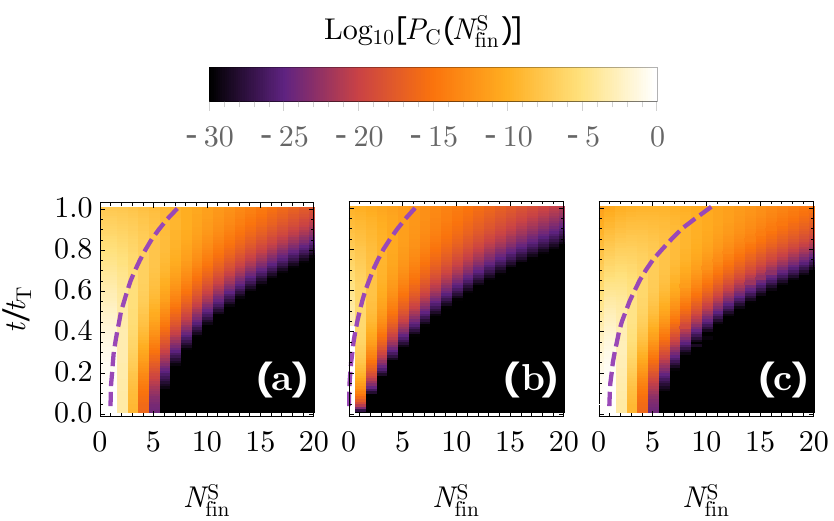}
    \caption{Time evolution inside the medium, from its input port ($t=0$) to the output port ($t=t_T$) of the probability distribution $P_\mathrm{C}$ to find $N^{\mathrm{S}}_{\mathrm{fin}}$ signal photons for three different initial bimodal coherent states $\ket{\alpha}_\mathrm{s}\ket{\beta}_\mathrm{i}$, (a)$\ket{1}_\mathrm{s}\ket{0}_\mathrm{i}$, (b) $\ket{0}_\mathrm{s}\ket{1}_\mathrm{i}$, and $\ket{1}_\mathrm{s}\ket{1}_\mathrm{i}$ calculated for $I_\mathrm{p}=0.2I_\mathrm{c}$. The dashed purple lines represent the time evolution of the expectation value $\braket{\hat{n}_\mathrm{s}}$.}
    \label{fig:PCoheVsTimeXalphaXbetaPlot}
\end{figure}

The time evolution of the probability distribution $P_\mathrm{C}$ is presented in figure (\ref{fig:PCoheVsTimeXalphaXbetaPlot}) for three different initial bimodal coherent states. In contrast with $P_\mathrm{F}$, this distribution is always single-peaked over the whole range of the interaction and its maximum shifts in time starting from $N^{\mathrm{S}}_{\mathrm{fin}}=|\alpha|^2$. It can also be noticed that, for a fixed $\alpha$, the distribution becomes wider and wider with the increase of $\beta$.

\section{Impedance matching, parameter space and noise performance}

To couple the JTWPA with its electromagnetic environment a characteristic impedance matching (e.g. $Z_\mathrm{c}=\SI{50}{\ohm}$) is commonly required. This target can be reached with non-trivial additional on-chip components or by properly tuning the cells parameters. To keep the induced magnetic flux function into the rf-SQUID single-valued, a design characterized by a screening parameter $\beta$, given by
\begin{align}\label{screening_parameter}
    \beta = \frac{2\pi L_\mathrm{g}I_\mathrm{c}}{\phi_0} < 1
\end{align}

is required. It is evident that a certain $\beta$ sets a hyperbolic relation between $L_\mathrm{g}$ and $I_\mathrm{c}$. Moreover, in standard fabrication techniques, the Josephson capacitance turns to be experimentally constrained to the critical current via a linear relation that links $C_\mathrm{J}$ with $I_\mathrm{c}$ passing through the junction area.\\
For completeness, it has to be noted that the proposed layout of JTWPA does not take into account any parasitic series inductances into the line. The presence of this stray circuit component tends to dilute the nonlinearity reducing the participation ratio of the Josephson nonlinearity \cite{manucharyan2007}; nonetheless this feature can be practically minimised by reducing the physical gap between two consecutive rf-SQUIDs, bringing this effect to be a small perturbation.\\
For a generic mode $n$, an expression for $C_\mathrm{g}$ having set $I_\mathrm{c}$ (consequently $L_\mathrm{g}$) and $Z_\mathrm{c}$ can be inferred starting from the relation for the characteristic impedance of a lossless transmission line ($Z_n=\sqrt{L_n / C_\mathrm{g}^n}$)
\begin{align}\label{ground_capacitance_constrain}
    C_\mathrm{g}^n &= \frac{L_n}{Z_n^2}
    = \frac{1}{Z_n^2}\frac{\Lambda_n L_\mathrm{g}}{1+\Lambda_n\frac{L_\mathrm{g}}{L_\mathrm{J}}} \nonumber \\
    &= \frac{\frac{L_\mathrm{g}}{1-L_\mathrm{g}C_\mathrm{J}\omega_n^2}}{Z_{n}^2\left( 1+ \frac{1}{1-L_\mathrm{g}C_\mathrm{J}\omega_n^2}\frac{L_\mathrm{g}}{\frac{\varphi_0}{I_\mathrm{c}}\frac{\Delta\Phi/\varphi_0}{\sin{\Delta\Phi/\varphi_0}}}\right)}
\end{align}

It has to be noticed that the impedance matching can be achieved just for a single-mode since the characteristic impedance $Z_n$ of the line is frequency dependant. The matched mode can be engineered \emph{ad-hoc} depending on the experiment requirements. If a low power reflection is required, the matched mode should be the pump one, instead, if no signal loss is preferred, the signal mode should be the matched one.
\begin{figure}[h] 
    \centering
    \includegraphics[width=9cm]{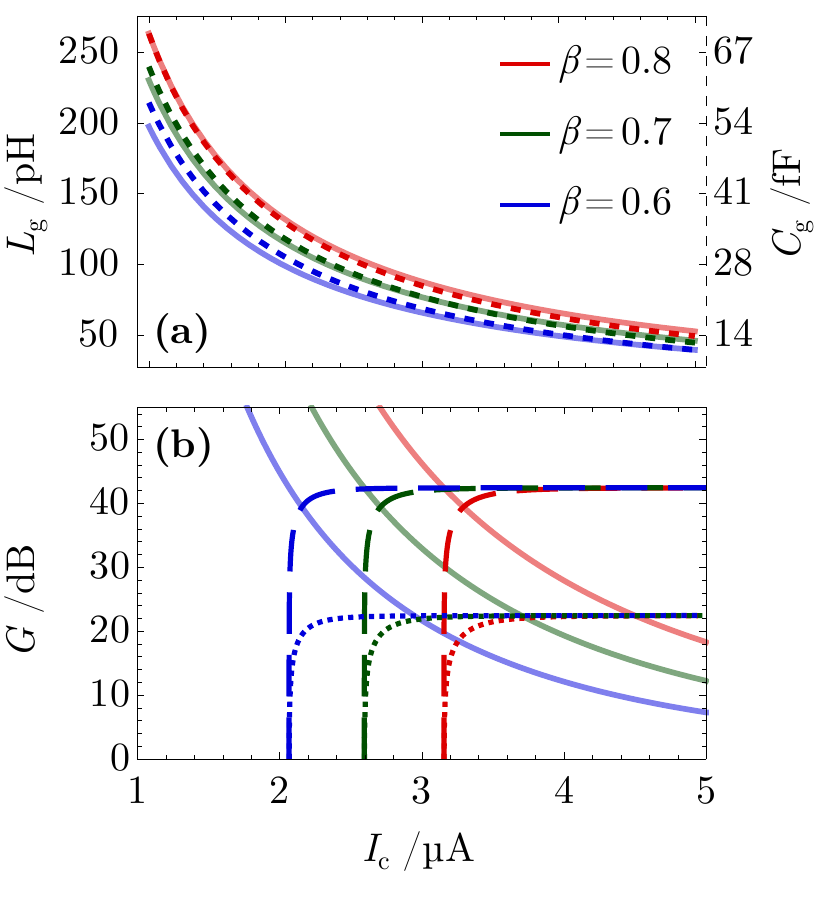}
    \caption{(a) The plot shows three sets of curves calculated for different values of the screening parameter $\beta$ representing the cell parameters for a $50 \: \Omega$ matching of the signal mode, at $\SI{7}{GHz}$, as a function of the critical current $I_\mathrm{c}$. The solid lines refer to the left axis and report the geometrical inductance $L_\mathrm{g}$ vs. $I_\mathrm{c}$. On the contrary, the dashed curves refer to the right axis and represent the ground capacitance $C_\mathrm{g}$ vs. $I_\mathrm{c}$. \\
    (b) Gain $G$ of the JTWPA in 3WM mode (solid line) as a function of the critical current $I_\mathrm{c}$ for different values of the screening parameter $\beta$. The dashed curves represent the limit above which the undepleted pump approximation cannot be considered valid considering the input state $\ket{1,0}$ while the dotted curves considering the input state $\ket{100,0}$ (see Supplemental Material Equation \eqref{eq:undepleted_gain_condition}).}
    \label{fig:ExpConstrainsPlot}
\end{figure}
Figure \ref{fig:ExpConstrainsPlot}(a) reports several curves representing the trends given by equations \eqref{screening_parameter} and \eqref{ground_capacitance_constrain} plotted as functions of $I_\mathrm{c}$ for different values of $\beta$ and for a $\SI{50}{\ohm}$ matching of a signal at $\SI{7}{GHz}$.\\
Figure \ref{fig:ExpConstrainsPlot} (b) shows instead the gain (solid lines) as a function of the Josephson critical current for different values of $\beta$. The dashed and dotted lines express the model validity limit corresponding to the undepleted pump approximation reported in Equation \ref{eq:undepleted_gain_condition}, considering $\ket{1,0}$ and $\ket{100,0}$ input states respectively and for each explored value of $\beta$. It is clear how the input photons number affects the maximum gain reached without pump depletion.

\section{Conclusions}

A quantum theory for parametric amplification via a chain of rf-SQUIDs embedded in a waveguide has been developed through a circuit-QED approach. A mixed lumped/distributed-element approach has been adopted to define the Hamiltonian of the system, valid for both 3WM and 4WM interactions. The dynamics of the system has been calculated first in the Heisenberg picture where, through the solution of a system of quantum Langevin equations for the traveling modes, a closed form for the evolution of the photonic populations, photon number gain and squeezing spectrum were found.
Then, using the interaction picture, the time evolution of some representative input states (Fock and coherent states) has been calculated, allowing to model the quantum dynamics of photonic amplification and virtual recombination into the JTWPA in the few photons regime.

\section*{Data Availability}

All data generated or analysed during this study are included in this published article.

\section*{Acknowledgments}

This research has been supported by DARTWARS, a project funded by Istituto Nazionale di Fisica Nucleare (INFN, National Scientific Committee 5), by the SUPERGALAX project in the framework of the European Union (EU) Horizon 2020 research and innovation programme (H2020, FETOPEN-2018-2020 call), and by the Joint Research Project PARAWAVE of the European Metrology Programme for Innovation and Research (EMPIR). This project (PARAWAVE) has received funding from the EMPIR programme co-financed by the Participating States and from the European Unions Horizon 2020 research and innovation programme.

\nocite{*}

\bibliography{paper}

\begin{acknowledgments}
This work has been partially funded by the SUPERGALAX project in the framework of the H2020-FETOPEN-2018-2020 call and the Joint Research Project PARAWAVE of the European Metrology Programme for Innovation and Research (EMPIR). This project has received funding from the EMPIR programme co-financed by the Participating States
and from the European Unions Horizon 2020 research and innovation programme.
\end{acknowledgments}

\section*{Author Contribution}
A.G., L.F., E.E. and L.C. conceptualized the work. A.G., L.F. and E.E. carried out the theoretical and numerical analysis. A.G., L.F. and E.E. wrote the manuscript. L.C. and A.M. participated in the discussion and editing of the manuscript.

\section*{Competing Interests}

The authors declare that there are no competing interests.
\clearpage

\pagebreak
\widetext
\begin{center}
\textbf{\large Supplemental Materials: A quantum model for rf-SQUIDs based metamaterials enabling 3WM and 4WM Traveling Wave Parametric Amplification}
\end{center}
\setcounter{equation}{0}
\setcounter{section}{0}
\setcounter{figure}{0}
\setcounter{table}{0}
\setcounter{page}{1}
\makeatletter
\renewcommand{\theequation}{S\arabic{equation}}
\renewcommand{\thefigure}{S\arabic{figure}}
\renewcommand{\thetable}{S\arabic{table}}
\renewcommand{\thesection}{S-\Roman{section}}
\renewcommand{\bibnumfmt}[1]{[S#1]}

\section{Experimental constrains and impedance matching}
\label{sub:50Ohm}

\subsection{Model validity due to Taylor expansions}
\label{sub:Validity_Taylor}
In subsection \ref{ch:Second quantization framework} two different Taylor expansions were performed regarding the Josephson energy into \eqref{eq:first_quantization_hamiltonian} and the nonlinear flux operator \eqref{ACflux}. In both expansions the phase swing $\delta \Phi$ is considered to be small, reflecting the amplitude of the AC current flowing into the transmission line. For this reason, it is necessary to point out the limit of validity of the model in terms of phase swing, hence of current. Since the pump current is considered to be much higher than the signal and idler ones (undepleted and classical pump approximations) it is legitimate to consider all the current flowing into the JTWPA equal to the pump current $I_\mathrm{p}$. This fact means that for the model to be valid, $I_\mathrm{p}$ needs to be smaller than a certain threshold.\\
An error function can be built for the Josephson energy and equation \eqref{ACflux} as the difference between the real function and its series expansion. The threshold is then chosen so that the error functions are always smaller than $5\%$, for any $I_\mathrm{p}$ used during the computations.

\begin{figure}[h]
    \centering
    \includegraphics[width=9 cm]{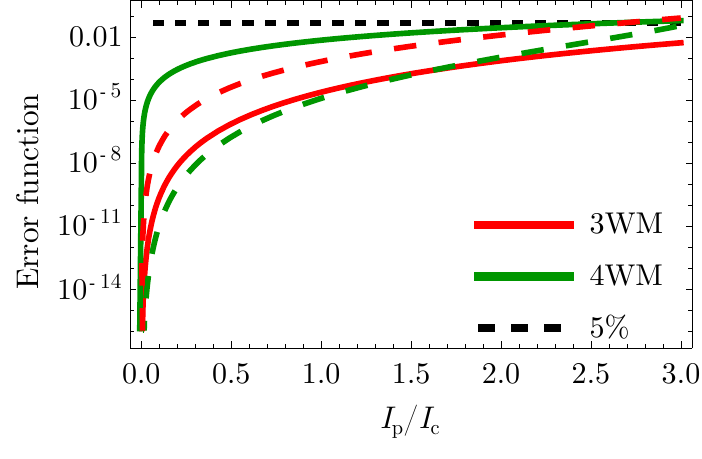}
    \caption{The solid curves represent the error function of \eqref{ACflux} while the dashed curves indicate the error function of the Josephson energy into \eqref{eq:first_quantization_hamiltonian}. The error functions are calculated as the relative difference between the function and its Taylor expansion versus the normalized AC pump ($I_\mathrm{p}/I_\mathrm{c}$) calculated in the 4WM and 3WM working points. The dashed black line indicates the $5\%$ threshold chosen as the reference error.   
    }\label{fig:SeriesExpErrorVsIp}
\end{figure}

Figure \ref{fig:SeriesExpErrorVsIp} shows the error functions calculated for equation \eqref{ACflux} (solid lines) and \eqref{eq:first_quantization_hamiltonian} (dashed lines) for $\Delta \Phi_{\mathrm{DC}}$ values corresponding to the 3WM and 4WM working points. This constraint fixes the maximum pump current allowed by the model. It's worth noting that the maximum allowed $I_\mathrm{p}$ value is greater than the Josephson critical current. This regime is accessible since the single cell is composed of the parallel of a geometric inductance and a Josephson junction, hence the total pump current can split unevenly between these two components, keeping the current flowing into the Josephson junction always below the critical value. It's worth mentioning that the resonances in the 4WM curves arise from the $I_\mathrm{p}$ values for which the Josephson inductance reaches its maximum value.

\section{Validity of the undepleted pump approximation}
\label{sub:Validity_Pump}
To obtain an analytical solution of the quantum Langevin equations the undepleted pump approximation has been made. Physically speaking this means that the pump power should be considered much higher than the signal and idler ones. This boundary directly translates into a simple relation for the number of pump photons into the amplifier, which must be always at least 10 times higher than the other modes.
\begin{align}\label{eq:undepleted_condition}
   \braket{\hat n_\mathrm{p}} > 10\cdot \braket{\hat n_{\mathrm{s}}}
\end{align}

Equation \eqref{eq:undepleted_condition} is a condition on the mode powers that has been recast using the number of photons.\\
Using \eqref{eq:linear_ampl_output} Equation \eqref{eq:undepleted_condition} becomes 
\begin{align}\label{eq:undepleted_gain_condition}
    \braket{\hat n_\mathrm{s}} &< \frac{\braket{\hat n_\mathrm{p}}}{10}\nonumber\\
    G\braket{\hat n_\mathrm{s,0}}+\braket{\mathcal{N}}&<\frac{\braket{\hat n_\mathrm{p}}}{10}\nonumber\\
    G&<\frac{\braket{\hat n_\mathrm{p}}}{10\braket{\hat n_\mathrm{s,0}}}-\frac{\braket{\mathcal{N}}}{\braket{\hat n_\mathrm{s,0}}}
\end{align}

the latter being a simple but powerful constraint on the maximum gain that the amplifier can show remaining well described by the undepleted pump approximation. This condition sets the limit at which the signal photonic population reaches the same order of magnitude as the pump one, hence the rate of annihilation of pump photons due to the generation of signal ones is no longer negligible. This limit is represented in Figure \ref{fig:ExpConstrainsPlot}(b) by the dashed curves for an input state $\ket{1,0}$ and by the dotted curves for an input state $\ket{100,0}$. For the screening parameter $\beta$ approaching unity the gain functions lose validity for higher values of the critical current because the non-linearity related to the induced flux into the rf-SQUID gets stronger. For lower values of $\beta$ the range of validity gets extended and the gain functions remain valid for lower critical currents.\\

\section{Hamiltonian linear density of the elementary cell components} \label{appendix:hamiltonian}

In this section we start describing the general method used to calculate the energy stored in a circuit element, then we derive the energy stored in each element which constitutes an elementary cell of a JTWPA.\\
Defining $I$ as the current flowing through a certain circuit element and $V$ as the voltage drop across it, the energy stored in the electrical component at a certain time $t$ can be expressed as the time-integrated power $P=VI$:
\begin{equation}
    U(t)=\displaystyle\int_{t_0}^t P(t')\mathrm{d}t'=\displaystyle\int_{t_0}^t I(t')\cdot V(t') \mathrm{d}t'
\end{equation}

The current flowing through a generic inductance $L$ induces a magnetic flux $\Delta\Phi(t)=LI(t)$, and can be related to the voltage drop across the element by the relation
\begin{equation}
    V(t)=L\frac{\mathrm{d}I(t)}{\mathrm{d}t}
\end{equation}

Hence one can express the energy stored in the geometrical inductance $L_\mathrm{g}$ as
\begin{align}
\label{eq:EC_Stored_Lg}
    U_{L_\mathrm{g}}(t)&=\displaystyle\int_{t_0}^t I_{L_\mathrm{g}}(t')\cdot V_{L_\mathrm{g}}(t') \mathrm{d}t'=\displaystyle\int_{t_0}^t I_{L_\mathrm{g}}(t')\cdot L_\mathrm{g}\frac{\mathrm{d}I_{L_\mathrm{g}}}{\mathrm{d}t'}\mathrm{d}t'\nonumber\\
    &=\frac{L_\mathrm{g}}{2}I^2_{L_\mathrm{g}}(t)=\frac{L_\mathrm{g}}{2}\left(\frac{\Delta\Phi(t)}{L_\mathrm{g}}\right)^2=\frac{\left(\Delta\Phi(t)\right)^2}{2L_\mathrm{g}}
\end{align}

having assumed $I_{L_\mathrm{g}}(t_0)=0$.\\
Exploiting the relation between magnetic flux difference and voltage drop
\begin{equation}
\label{eq:rel_Voltage_Flux}
V(t)=\frac{\mathrm{d}\Delta\Phi(t)}{\mathrm{d}t}
\end{equation}

the energy stored in the nonlinear Josephson inductance $L_\mathrm{J}$ can be expressed as
\begin{align}
    U_{L_\mathrm{J}}(t)&=\displaystyle\int_{t_0}^t I_{L_\mathrm{J}}(t')\cdot V_{L_\mathrm{J}}(t')\mathrm{d}t'\nonumber\\
    &=\displaystyle\int_{t_0}^t I_\mathrm{c} \sin{\left(\frac{\Delta\Phi(t')}{\varphi_0}\right)}\cdot \frac{\mathrm{d}\Delta\Phi(t')}{\mathrm{d}t'}\mathrm{d}t'\nonumber\\
    &=\varphi_0 I_\mathrm{c} \left(1-\cos{\left(\frac{\Delta\Phi(t)}{\varphi_0}\right)}\right)
\end{align}

having assumed $\Delta\Phi(t_0)=0$.\\
Exploiting the relation between the current flowing through a capacitance $C$ and the voltage drop $V$ across its terminals
\begin{equation}
\label{eq:rel_IV_Capacitance}
    I(t)=C\frac{\mathrm{d}V(t)}{\mathrm{d}t}
\end{equation}

the energy stored in the ground capacitance $C_\mathrm{g}$ can be expressed as
\begin{align} \label{capacitanceenergyvoltage}
U_{C_\mathrm{g}}(t)&=\displaystyle\int_{t_0}^t I_{C_\mathrm{g}}(t')\cdot V_{C_\mathrm{g}}(t')\mathrm{d}t'\nonumber\\
&=\displaystyle\int_{t_0}^t C_\mathrm{g} \frac{\mathrm{d}V_{C_\mathrm{g}}(t')}{\mathrm{d}t'}\cdot V_{C_\mathrm{g}}(t')\mathrm{d}t'\nonumber\\
&=\frac{C_\mathrm{g}}{2}V^2_{C_\mathrm{g}}(t)= \frac{1}{2C_\mathrm{g}}Q^2_{C_\mathrm{g}}
\end{align}

having assumed $V_{C_\mathrm{g}}(t_0)=0$.\\
Lastly, exploiting relations \eqref{eq:rel_Voltage_Flux} and \eqref{eq:rel_IV_Capacitance}, the energy stored in the capacitance associated with the Josephson junction can be expressed as
\begin{align}
    U_{C_\mathrm{J}}(t)&=\displaystyle\int_{t_0}^tI_{C_\mathrm{J}}(t')\cdot V_{C_\mathrm{J}}(t')\mathrm{d}t'\nonumber\\
    &=C_\mathrm{J}\displaystyle\int_{t_0}^t\frac{\mathrm{d}}{\mathrm{d}t'}\left[\frac{\mathrm{d}\Delta\Phi(t')}{\mathrm{d}t'}\right]\cdot \frac{\mathrm{d}\Delta\Phi(t')}{\mathrm{d}t'}\mathrm{d}t'\nonumber\\
    &=\frac{C_\mathrm{J}}{2}\left(\frac{\mathrm{d}\Delta\Phi(t)}{\mathrm{d}t}\right)^2
\end{align}
having assumed $\Delta\Phi(t_0)=0$.\\

Using a standard procedure \cite{Vool2017b}, one can derive the Hamiltonian operator that describes an electrical circuit starting from the definition of the energy stored in each of its components and transforming the physical observables into the corresponding operators. Furthermore, being $a$ as the unit cell length, one can express the linear density of Hamiltonian associated with each component of the circuit represented in Figure \ref{fig:RfSquid_Array} as
\begin{equation}
    \mathcal{\hat{H}}_{L_\mathrm{g}}=\frac{1}{2aL_\mathrm{g}}\Delta\hat{\Phi}^2
\end{equation}
\begin{equation}
    \mathcal{\hat{H}}_{L_\mathrm{J}}=\frac{\varphi_0I_\mathrm{c}}{a}\left(1-\cos{\left(\frac{\Delta \hat{\Phi}}{\varphi_0}\right)}\right)
\end{equation}
\begin{equation}
    \mathcal{\hat{H}}_{C_\mathrm{g}}=\frac{C_\mathrm{g}}{2a}\hat{V}^2_{C_\mathrm{g}}=\frac{1}{2aC_\mathrm{g}}\hat{Q}^2_{C_\mathrm{g}}
\end{equation}
\begin{equation}
    \mathcal{\hat{H}}_{C_\mathrm{J}}=\frac{C_\mathrm{J}}{2a}\left(\frac{\partial \Delta\hat{\Phi}}{\partial t}\right)^2
\end{equation}

\section{Inductance of the unit cell} \label{appendix:effectiveinductance}

It is useful to define an effective inductance that takes into account the parallel effect of the geometric inductance $L_\mathrm{g}$ and the Josephson capacitance $C_\mathrm{J}$. Keeping in mind that the impedance of an inductor $L$ for the mode $n$ is $Z_L = j\omega_n L$ while the impedance of a capacitor $C$ for the same mode is $Z_\mathrm{c} = 1/j\omega_n C$, we can write
\begin{align} 
\frac{1}{Z_{L_{\mathrm{eff},n}}} &=  \frac{1}{Z_{L_\mathrm{g}}} + \frac{1}{Z_{C_\mathrm{J}}} \nonumber\\ 
\frac{1}{j\omega_n L_{\mathrm{eff},n}} &=  \frac{1}{j\omega_n L_\mathrm{g}} + j\omega C_\mathrm{J} \nonumber\\
\frac{1}{L_{\mathrm{eff},n}} &=  \frac{1}{L_\mathrm{g}} - \omega_n^2 C_\mathrm{J} =  \frac{1 - \omega_n^2 L_\mathrm{g} C_\mathrm{J}}{L_\mathrm{g}}
\end{align}

It is found that the dispersion coefficient of the $n$-th mode ($\Lambda_n$) can be defined by the relation
\begin{align}\label{effective_ind}
L_{\mathrm{eff},n} &= \frac{L_\mathrm{g}}{1 - \omega_n^2 L_\mathrm{g} C_\mathrm{J}} \equiv \Lambda_n L_\mathrm{g}
\end{align}

It is now possible to compute the total inductance of the elementary cell by calculating the parallel of the effective inductance $L_{\mathrm{eff},n}$ and the Josephson inductance $ L_\mathrm{J}$
\begin{equation} \label{eq12}
\begin{split}
\frac{1}{L_n}
& = \frac{1}{L_\mathrm{J}} + \frac{1}{L_{\mathrm{eff},n}}=\frac{L_\mathrm{J} L_{\mathrm{eff},n}}{L_\mathrm{J} + L_{\mathrm{eff},n}}
\end{split}
\end{equation}

Hence using equations \eqref{effective_ind} and \eqref{eq:Lj} the unit cell inductance can be written as
\begin{align}\label{eq13}
L_n
& = \frac{L_{\mathrm{J}_0}L_\mathrm{g}\left(\Delta\Phi/\varphi_{0}\right)}{L_{\mathrm{J}_0}\left(\Delta\Phi/\varphi_{0}\right)\left(\frac{L_\mathrm{g}}{L_{\mathrm{J}_0}}\frac{\sin{\left(\Delta\Phi/\varphi_{0}\right)}}{\left(\Delta\Phi/\varphi_{0}\right)}+1-L_\mathrm{g}C_\mathrm{J}\omega_n^2\right)}\nonumber\\
& = \frac{L_\mathrm{g}}{\frac{L_\mathrm{g}}{L_{\mathrm{J}_0}}\frac{\sin{\left(\Delta\Phi/\varphi_{0}\right)}}{\left(\Delta\Phi/\varphi_{0}\right)}+1-L_\mathrm{g}C_\mathrm{J}\omega_n^2}\nonumber\\
&=\frac{\Lambda_{n}L_\mathrm{g}}{1+\Lambda_{n}\frac{L_\mathrm{g}}{L_{\mathrm{J}_0}}\frac{\sin{\left(\Delta\Phi/\varphi_{0}\right)}}{\left(\Delta\Phi/\varphi_{0}\right)}}
\end{align}

\section{Time-dependent flux operator}\label{appendix:nonlinear_flux}
The time-dependent component of the flux difference operator can be found through the constitutive equation of an inductor \eqref{eq4modes}, where a mode decomposition has been performed. It has to be noticed that the inductance can be mode-dependent. The current through the unit cell can be calculated exploiting the telegrapher's equation \eqref{eq10}, where the voltage drop to ground comes from \eqref{eq9} and the cell inductance of the $n$-th mode is found from equation \eqref{eq13}. Hence, the AC current passing through the line results to be
\begin{equation}\label{eq11}
    \hat I(z,t)= \sum_n sgn(n) \sqrt{\frac{\hbar\omega_n}{2\hat{L}_{n}N}}\left(\hat a_{n}e^{i(k_nz-\omega_nt)}+\mathrm{H.c.}\right)
\end{equation}

the nonlinear time-dependent flux operator $\delta \hat \Phi$ is then obtained exploiting \eqref{eq13} by replacing the classical variable $\Delta \Phi$ by corresponding operator $\Delta \hat{\Phi}$
\begin{align} \label{eq14}
\delta\hat{\Phi}
& =\sum_n sgn(n) \sqrt{\frac{\hbar\omega_n}{2L_nN}}L_n\left(\hat a_{n}e^{i(k_nz-\omega_nt)}+\mathrm{H.c.} \right)\nonumber\\
& =\sum_n sgn(n) \sqrt{\frac{\hbar\omega_n}{2N}}\sqrt{\Lambda_{n}L_\mathrm{g}}\cdot\nonumber\\
&\hspace{1cm}\cdot\Bigg(1+\Lambda_{n}\frac{L_\mathrm{g}}{L_{\mathrm{J}_0}}\frac{\sin{\frac{\Delta\Phi_{\mathrm{DC}} + \delta\hat{\Phi}}{\varphi_{0}}}}{\frac{\Delta\Phi_{\mathrm{DC}} + \delta\hat{\Phi}}{\varphi_{0}}}\Bigg)^{-1/2}\cdot\nonumber\\
&\hspace{1cm}\cdot\left(\hat a_{n}e^{i(k_nz-\omega_nt)}+\mathrm{H.c.}\right)=\nonumber\\
& =\sum_n \left(1+\Lambda_{n}\frac{L_\mathrm{g}}{L_{\mathrm{J}_0}}\frac{\sin{\frac{\Delta\Phi_{\mathrm{DC}} + \delta\hat{\Phi}}{\varphi_{0}}}}{\frac{\Delta\Phi_{\mathrm{DC}} + \delta\hat{\Phi}}{\varphi_{0}}}\right)^{-1/2}\delta\hat{\Phi}^{(0)}_{n}
\end{align}

where we have identified
\begin{align}
\label{eq:DeltaPhi0_n}
    \delta\hat{\Phi}^{(0)}_{n} &\equiv sgn(n) \sqrt{\frac{\hbar\omega_n}{2N}}\sqrt{\Lambda_{n}L_\mathrm{g}}\left(\hat a_{n}e^{i(k_nz-\omega_nt)}+\mathrm{H.c.}\right)\nonumber\\
   &= c_n\left(\hat a_{n}e^{i(k_nz-\omega_nt)}+\mathrm{H.c.}\right)
\end{align}

with
\begin{equation}\nonumber
    c_{n} = sgn(n)\sqrt{\frac{\hbar\omega_n}{2N}}\sqrt{L_\mathrm{g}\Lambda_{n}}
\end{equation}

Equation \eqref{eq14} is a implicit relation that involves $\delta\hat{\Phi}$ and can be solved at zero order by the substitution $\delta\hat{\Phi} \mapsto \delta\hat{\Phi}^{(0)}=\sum_n\delta\hat{\Phi}^{(0)}_{n}$ on the right hand side, so that we get
\begin{equation}\label{eq16}
    \delta\hat{\Phi} =\sum_n\left[ \left(1+\Lambda_{n}\frac{L_\mathrm{g}}{L_{\mathrm{J}_0}}\frac{\sin{\frac{\Delta\Phi_{\mathrm{DC}} + \delta\hat{\Phi}^{(0)}}{\varphi_{0}}}}{\frac{\Delta\Phi_{\mathrm{DC}} + \delta\hat{\Phi}^{(0)}}{\varphi_{0}}}\right)^{-1/2}\delta\hat{\Phi}^{(0)}_{n}\right]
\end{equation}

By invoking the Taylor expansion of the square root into equation \eqref{eq16} for $\delta\hat{\Phi}^{(0)}\ll \varphi_0$, one obtains
\begin{align}\label{eq17}
    \delta\hat{\Phi}&=\sum_n\Bigg[q_{0,n}+q_{1,n}\left(\delta\hat{\Phi}^{(0)}\right)+q_{2,n}\left(\delta\hat{\Phi}^{(0)}\right)^2+\nonumber\\
    &+q_{3,n}\left(\delta\hat{\Phi}^{(0)}\right)^3+O\left(\delta\hat{\Phi}^{(0)}\right)^4
    \Bigg] \delta\hat{\Phi}^{(0)}_{n}
\end{align}

We stress that the terms $q_{0,n}$, $q_{1,n}$, $q_{2,n}$ and $q_{3,n}$ are coefficients of a Taylor expansion and result to be functions of the external bias conditions (i.e., of the constant flux difference $\Delta\Phi_{\mathrm{DC}}$).
It's worth noting here how the lowest perturbative order approach adopted in equation \eqref{eq16} takes into account interactions of modes at the first order, which means a single multimode interaction. While the power expansion truncation up to the third order in equation \eqref{eq17} limits our model to the interaction of a single-mode ($\delta\hat{\Phi}_n^{(0)}$) with up to three modes. For a quantitative comparison of the power expansion approach \eqref{eq17} respect to the bare nonlinearity \eqref{eq16} see Figure \ref{fig:SeriesExpErrorVsIp}.

\section{Coupling Coefficients}
\label{app:CouplingCoefficients}
Defined $p_1=\cos{\left(\Delta\Phi_{\mathrm{DC}}/\varphi_{0}\right)}$ and $p_2=\sin{\left(\Delta\Phi_{\mathrm{DC}}/\varphi_{0}\right)}$ the analytical forms of the coupling coefficient in the 3WM and 4WM Hamiltonians \eqref{eq:H3WM_Full}-\eqref{eq:H4WM_Full} are:

\begin{align}\label{eq:h0}
    \chi_0&=\frac{N}{\hbar}\left[I_\mathrm{c}\varphi_0\left(1-\cos{\left(\frac{\Delta\Phi_{\mathrm{DC}}}{\varphi_0}\right)}\right)+\frac{\Delta\Phi^2_{\mathrm{DC}}}{2L_\mathrm{g}}\right]
\end{align}

\begin{align}\label{eq:xin}
    \chi_1^{(n)}&=\frac{\omega_n}{2}\left(1+2L_\mathrm{g}\Lambda_n \left[\left(I_\mathrm{c}p_2+\frac{\Delta\Phi_{\mathrm{DC}}}{L_\mathrm{g}}\right)q_{1,n}+\left(\frac{I_\mathrm{c}p_1}{\varphi_0}+\frac{1}{L_\mathrm{g}}+C_\mathrm{J}\Delta\omega_n^2\right)\frac{q_{0,n}^2}{2}\right]\right)
\end{align}

\begin{align}
\label{eq:chi3_nlm}
\chi_3^{(n,l,m)}&=\sqrt{\frac{\hbar L_\mathrm{g}^3}{8N}}\sqrt{\omega_n\Lambda_n\omega_l\Lambda_l\omega_m\Lambda_m}\Bigg[\left(I_\mathrm{c}p_2+\frac{\Delta\Phi_{\mathrm{DC}}}{L_\mathrm{g}}\right)q_{2,n}+\left(\frac{I_\mathrm{c}p_1}{\varphi_0}+\frac{1}{L_\mathrm{g}}\right)q_{0,n}q_{1,l}+\nonumber\\
    &\hspace{1.5cm}-\frac{I_\mathrm{c}p_2}{6\varphi_0^2}q_{0,n}q_{0,l}q_{0,m}+\frac{C_\mathrm{J}}{2}\left[q_{0,n}q_{1,l}\;\Delta\omega_n\Delta\omega_{m,l}+q_{1,n}q_{0,l}\;\Delta\omega_l\Delta\omega_{n,m}\right]\Bigg]
\end{align}

\begin{align}
\label{eq:chi4_nlms}
\chi_4^{(n,l,m,s)}&=\frac{\hbar L_\mathrm{g}^2}{4N}\sqrt{\omega_n\Lambda_n\omega_l\Lambda_l\omega_m\Lambda_m\omega_s\Lambda_s}\Bigg[\left(I_\mathrm{c}p_2+\frac{\Delta\Phi_{\mathrm{DC}}}{L_\mathrm{g}}\right)q_{3,n}+\frac{1}{2}\left(\frac{I_\mathrm{c}p_1}{\varphi_0}+\frac{1}{L_\mathrm{g}}\right)\left(2q_{0,n}q_{2,l}+q_{1,n}q_{1,l}\right)+\nonumber\\
&-\frac{I_\mathrm{c}p_2}{2\varphi^2_0}\;q_{1,n}q_{0,l}q_{0,m} -\frac{I_\mathrm{c}p_1}{24\varphi^3_0}\;q_{0,n}q_{0,l}q_{0,m}q_{0,s}+ \nonumber \\
&+\frac{C_\mathrm{J}}{2}\left[q_{1,n}q_{1,l}\left(\Delta\omega_m\Delta\omega_{s,l}+\Delta\omega_n\Delta\omega_{m,l}\right)+q_{2,n}q_{0,l}\;\Delta\omega_l\Delta\omega_{2m,n}+q_{0,n}q_{2,l}\;\Delta\omega_n\Delta\omega_{2m,l}\right] \Bigg]
\end{align}

\section{Time evolution of the ladder operators: coupled mode equations}\label{appendix:coupled_mode_equations}

Starting from equation \eqref{eq:H3WM_Full} and \eqref{eq:H4WM_Full} it is possible to work out the dynamics of the system in 3WM and 4WM regime.\\
In 3WM regime, hence using \eqref{eq:H3WM_Full} to compute the Heisenberg equation it is obtained
\begin{align}\label{eq:dap_on_dt_3WM}
    \frac{\mathrm{d}\hat{a}_\mathrm{p}}{\mathrm{d}t}
    &=\frac{i}{\hbar} \left[\hat{H}^{\{\mathrm{p,s,i}\}}_{\mathrm{3WM}},\hat{a}_\mathrm{p}\right]= -i\left[ \chi_1^\mathrm{p} \hat A_\mathrm{p} +  \chi_3^{\{\mathrm{p,s,i}\}} \hat{a}_\mathrm{s} \hat{a}_\mathrm{i} \right]
\end{align}

\begin{align}\label{eq:das_on_dt_3WM}
    \frac{\mathrm{d}\hat{a}_\mathrm{s}}{\mathrm{d}t}
    &=\frac{i}{\hbar} \left[\hat{H}^{\{\mathrm{p,s,i}\}}_{\mathrm{3WM}},\hat{a}_\mathrm{s}\right]= -i\left[ \chi_1^\mathrm{s} \hat{a}_\mathrm{s} +  \chi_3^{\{\mathrm{p,s,i}\}} \hat A_\mathrm{p} \hat{a}_\mathrm{i}^\dagger \right]
\end{align}

\begin{align}\label{eq:dai_on_dt_3WM}
    \frac{\mathrm{d}\hat{a}_\mathrm{i}}{\mathrm{d}t}
    &=\frac{i}{\hbar} \left[\hat{H}^{\{\mathrm{p,s,i}\}}_{\mathrm{3WM}},\hat{a}_\mathrm{i}\right]= -i\left[ \chi_1^\mathrm{i} \hat{a}_\mathrm{i} +  \chi_3^{\{\mathrm{p,s,i}\}} \hat A_\mathrm{p} \hat{a}_\mathrm{s}^\dagger \right]
\end{align}

While in 4WM regime, through \eqref{eq:H4WM_Full}
\begin{align}\label{eq:dap_on_dt_4WM}
    \frac{\mathrm{d}\hat{a}_\mathrm{p}}{\mathrm{d}t}
    &=\frac{i}{\hbar} \left[\hat{H}^{\{\mathrm{p,s,j}\}}_{\mathrm{4WM}},\hat{a}_\mathrm{p}\right]\nonumber \\
    &= -i\Big[ \Big( \xi_\mathrm{p}+\xi_{\mathrm{pp}}+2\xi_{\mathrm{pp}}\hat A_\mathrm{p}^{\dagger}\hat A_\mathrm{p} +\nonumber\\
    &+ \xi_{\mathrm{ps}}\hat{a}_\mathrm{s}^{\dagger}\hat A_\mathrm{p} +\xi_\mathrm{pj}\hat{a}_\mathrm{j}^{\dagger}\hat{a}_\mathrm{j}\Big)\hat A_\mathrm{p} +  2\chi_4^{\{\mathrm{p,p,s,i}\}}\hat A_\mathrm{p}^\dagger \hat{a}_\mathrm{s} \hat{a}_\mathrm{j} \Big]
\end{align}

\begin{align}\label{eq:das_on_dt_4WM}
     \frac{\mathrm{d}\hat{a}_\mathrm{s}}{\mathrm{d}t}
    &=\frac{i}{\hbar} \left[\hat{H}^{\{\mathrm{p,s,i}\}}_{\mathrm{4WM}},\hat{a}_\mathrm{s}\right]\nonumber \\
    &= -i\Big[ \Big( \xi_\mathrm{s}+\xi_\mathrm{ss}+2\xi_\mathrm{ss}\hat{a}_\mathrm{s}^{\dagger}\hat{a}_\mathrm{s} + \xi_{\mathrm{ps}}\hat A_\mathrm{p}^{\dagger}\hat A_\mathrm{p} + \nonumber\\
    & +\xi_\mathrm{sj}\hat{a}_\mathrm{j}^{\dagger}\hat{a}_\mathrm{j}\Big)\hat{a}_\mathrm{s} +  \chi_4^{\{\mathrm{p,p,s,i}\}}\hat A_\mathrm{p} \hat A_\mathrm{p} \hat{a}_\mathrm{j}^\dagger \Big]
\end{align}

\begin{align}\label{eq:dai_on_dt_4WM}
     \frac{\mathrm{d}\hat{a}_\mathrm{i}}{\mathrm{d}t}
    &=\frac{i}{\hbar} \left[\hat{H}^{\{\mathrm{p,s,i}\}}_{\mathrm{4WM}},\hat{a}_\mathrm{i}\right]\nonumber \\
    &= -i\Big[ \Big( \xi_\mathrm{j}+\xi_\mathrm{jj}+2\xi_\mathrm{jj}\hat{a}_\mathrm{i}^{\dagger}\hat{a}_\mathrm{j} + \xi_\mathrm{pj}\hat A_\mathrm{p}^{\dagger}\hat A_\mathrm{p} + \nonumber \\
    &+ \xi_\mathrm{sj}\hat{a}_\mathrm{s}^{\dagger}\hat{a}_\mathrm{s}\Big)\hat{a}_\mathrm{i} +  \chi_4^{\{\mathrm{p,p,s,j}\}}\hat A_\mathrm{p} \hat A_\mathrm{p} \hat{a}_\mathrm{s}^\dagger \Big]
\end{align}

The systems composed by equations \eqref{eq:dap_on_dt_3WM}, \eqref{eq:das_on_dt_3WM}, \eqref{eq:dai_on_dt_3WM} and by equations \eqref{eq:dap_on_dt_4WM}, \eqref{eq:das_on_dt_4WM}, \eqref{eq:dai_on_dt_4WM} are known as quantum Langevin equations (Coupled Mode Equations in the classical regime), and their solutions determine the time evolution of the modes interacting into the JTWPA. The undepleted pump approximation describes a regime where the signal and idler modes can be considered small compared to the pump mode. This approximation allows to solve analytically the system of coupled differential equations by substituting the ladder operator of the pump mode with its classical counterpart, having defined
\begin{align}\label{classical_pump}
    \sqrt{\frac{2\hbar\omega_\mathrm{p}}{C_\mathrm{g}N}}\hat{a}_\mathrm{p} \mapsto A_\mathrm{p}
\end{align}

as the classical voltage amplitude of Equation \eqref{eq9}.\\
In the 4M case, equation \eqref{classical_pump} can be substituted into \eqref{eq:dap_on_dt_4WM}
\begin{align}\label{eq:CMEdAp_on_dt_4WM}
    \sqrt{\frac{C_\mathrm{g}N}{2\hbar\omega_\mathrm{p}}}\frac{\mathrm{d}A_\mathrm{p}}{\mathrm{d}t}&= -i\Bigg[ \left( \xi_\mathrm{p}+\xi_{\mathrm{pp}} \right)\sqrt{\frac{C_\mathrm{g}N}{2\hbar\omega_\mathrm{p}}}A_\mathrm{p} +\nonumber\\
    & + 2\left(\frac{C_\mathrm{g}N}{2\hbar \omega_\mathrm{p}}\right)^{\frac{3}{2}}|A_\mathrm{p}|^2\xi_{\mathrm{pp}}A_\mathrm{p}+\xi_{\mathrm{ps}}\sqrt{\frac{C_\mathrm{g}N}{2\hbar\omega_\mathrm{p}}}A_\mathrm{p}\hat{a}_\mathrm{s}^\dagger \hat{a}_\mathrm{s} +\nonumber\\
    & + \xi_\mathrm{pj}\sqrt{\frac{C_\mathrm{g}N}{2\hbar\omega_\mathrm{p}}}\hat{a}_\mathrm{j}^\dagger \hat{a}_\mathrm{j} A_\mathrm{p}  + 2\chi_4^{\{\mathrm{p,p,s,j}\}}\sqrt{\frac{C_\mathrm{g}N}{2\hbar\omega_\mathrm{p}}}A_\mathrm{p}^*\hat{a}_\mathrm{s} \hat{a}_\mathrm{j} \Bigg]\nonumber\\
    \frac{\mathrm{d}A_\mathrm{p}}{\mathrm{d}t}&=-i\Bigg[ \left( \xi_\mathrm{p}+\xi_{\mathrm{pp}} \right)A_\mathrm{p} + 2\frac{C_\mathrm{g}N}{2\hbar\omega_\mathrm{p}}|A_\mathrm{p}|^2\xi_{\mathrm{pp}}A_\mathrm{p}+\nonumber\\
    &+\xi_{\mathrm{ps}}A_\mathrm{p}\hat{a}_\mathrm{s}^\dagger \hat{a}_\mathrm{s} + \xi_\mathrm{pj}\hat{a}_\mathrm{j}^\dagger \hat{a}_\mathrm{j} A_\mathrm{p} + 2\chi_4^{p,p,s,j}A_\mathrm{p}^*\hat{a}_\mathrm{s} \hat{a}_\mathrm{j} \Bigg]
\end{align}

Keeping the leading terms in \eqref{eq:CMEdAp_on_dt_4WM} we get
\begin{align}\label{eq:CMEdAp_on_dt4WM}
    \frac{\mathrm{d}A_\mathrm{p}}{\mathrm{d}t}&\approx-i\left[ \left( \xi_\mathrm{p}+\xi_{\mathrm{pp}} \right)A_\mathrm{p} + 2\frac{C_\mathrm{g}N}{2\hbar\omega_\mathrm{p}}|A_\mathrm{p}|^2\xi_{\mathrm{pp}}A_\mathrm{p} \right]\nonumber\\
    &= -i\left( \left( \xi_\mathrm{p}+\xi_{\mathrm{pp}} \right) + 2\frac{C_\mathrm{g}N}{2\hbar\omega_\mathrm{p}}|A_\mathrm{p}|^2\xi_{\mathrm{pp}} \right)A_\mathrm{p} \nonumber\\
    &=-i\Psi_\mathrm{p}A_\mathrm{p}
\end{align}

and by solving this latter, one can derive
\begin{equation}\label{eq:Apt4WM}
    A_\mathrm{p}(t) = |A_{\mathrm{p},0}|e^{-i\Psi_\mathrm{p} t}
\end{equation}

with
\begin{equation}
    \Psi_\mathrm{p}= \xi_\mathrm{p}+\xi_{\mathrm{pp}} + 2\xi_{\mathrm{pp}}\frac{C_\mathrm{g}N}{2\hbar\omega_\mathrm{p}}|A_\mathrm{p}|^2
\end{equation}

$|A_{\mathrm{p},0}|$ is the voltage amplitude at $t=0$, the time in which the mode enters in the non-linear medium. For sake of simplicity, we have assumed the initial phase of $A_\mathrm{p}$ equal to zero. Similarly, the time evolution for the signal and idler annihilation operators can be written as
\begin{align}\label{eq:das_on_dt4WM_ClassicalPump}
    \frac{\mathrm{d}\hat{a}_\mathrm{s}}{\mathrm{d}t}&= -i\Bigg[ \Big( \xi_\mathrm{s}+\xi_\mathrm{ss}+2\xi_\mathrm{ss}\hat{a}_\mathrm{s}^{\dagger}\hat{a}_\mathrm{s} + \xi_{\mathrm{ps}} \left( \frac{C_\mathrm{g}N}{2\hbar\omega_\mathrm{p}} \right)|A_\mathrm{p}|^2 +\nonumber\\
    &+\xi_{si}\hat{a}_\mathrm{j}^{\dagger}\hat{a}_\mathrm{j}\Big)\hat{a}_\mathrm{s} +  \chi_4^{\{\mathrm{p,p,s,j}\}}\left( \frac{C_\mathrm{g}N}{2\hbar\omega_\mathrm{p}} \right)A_\mathrm{p}^2 \hat{a}_\mathrm{j}^\dagger \Bigg]\nonumber\\
    &\approx -i\Bigg[ \left( \xi_\mathrm{s}+ \xi_{\mathrm{ps}} \left( \frac{C_\mathrm{g}N}{2\hbar\omega_\mathrm{p}} \right)|A_\mathrm{p}|^2 \right)\hat{a}_\mathrm{s} +\nonumber\\
    &+\chi_4^{\{\mathrm{p,p,s,j}\}}\left( \frac{C_\mathrm{g}N}{2\hbar\omega_\mathrm{p}} \right)A_\mathrm{p}^2 \hat{a}_\mathrm{j}^\dagger\Bigg]\nonumber\\
    &=-i\Bigg[ \Psi_\mathrm{s} \hat{a}_\mathrm{s} +  \chi_4^{\{\mathrm{p,p,s,j}\}}\left( \frac{C_\mathrm{g}N}{2\hbar\omega_\mathrm{p}} \right)A_\mathrm{p}^2 \hat{a}_\mathrm{j}^\dagger\Bigg]
\end{align}

with
\begin{align}
    \Psi_\mathrm{s}=\xi_\mathrm{s} + \xi_{\mathrm{ps}} \left( \frac{C_\mathrm{g}N}{2\hbar\omega_\mathrm{p}} \right)|A_\mathrm{p}|^2
\end{align}

In the co-rotating frame
\begin{align}\label{eq:das_on_dt4WM_ClassicalPumP_Corotating}
    \frac{\mathrm{d}\hat{a}_\mathrm{s}}{\mathrm{d}t}&= 
    -i \chi_4^{\{\mathrm{p,p,s,j}\}}\left( \frac{C_\mathrm{g}N}{2\hbar\omega_\mathrm{p}} \right)A_\mathrm{p}^2 \left(\hat{a}_\mathrm{j}^{CR}\right)^\dagger e^{i(\Psi_\mathrm{s}+\Psi_\mathrm{j})t}\nonumber\\
    &=-i\chi_4|A_{\mathrm{p},0}|^2\left(\hat{a}_\mathrm{j}^{CR}\right)^\dagger e^{-i(2\Psi_\mathrm{p}-\Psi_\mathrm{s}-\Psi_\mathrm{j})t}\nonumber\\
    &=-i\chi_4|A_{\mathrm{p},0}|^2\left(\hat{a}_\mathrm{j}^{CR}\right)^\dagger e^{-i\Psi_4 t}
\end{align}
where equation \eqref{eq:Apt4WM} has been exploited, having
\begin{align}\label{eq:phase_mismatch_density_4WM}
    \Psi_4 = 2\Psi_\mathrm{p} -\Psi_\mathrm{s} - \Psi_\mathrm{j}
\end{align}
and  introducing
\begin{align}\label{eq_chi4}
    \chi_4=\chi_4^{\{\mathrm{p,p,s,j}\}} \frac{C_\mathrm{g}N}{2\hbar\omega_\mathrm{p}}
\end{align}

The 3WM system can be solved through the same procedure starting from equation \eqref{eq:dap_on_dt_3WM}
\begin{align}
    \frac{\mathrm{d}\hat{a}_\mathrm{p}}{\mathrm{d}t}
    &=\frac{i}{\hbar} \left[\hat{H}^{\{\mathrm{p,s,i}\}}_{\mathrm{3WM}},\hat{a}_\mathrm{p}\right]\nonumber\\
    \frac{\mathrm{d}A_\mathrm{p}}{\mathrm{d}t}&= -i\left[ \chi_1^\mathrm{p} \hat A_\mathrm{p} + \chi_3^{\{\mathrm{p,s,i}\}} \hat{a}_\mathrm{s} \hat{a}_\mathrm{i} \right] \approx-i\chi_1^\mathrm{p} A_\mathrm{p} 
\end{align}

whose solution is
\begin{equation}
    A_\mathrm{p}(t) = |A_{\mathrm{p},0}|e^{-i\chi_1^\mathrm{p} t}
\end{equation}

equation \eqref{eq:das_on_dt_3WM} becomes
\begin{align}
    \frac{\mathrm{d}\hat{a}_\mathrm{s}}{\mathrm{d}t}
    &= -i\left[ \chi_1^\mathrm{s} \hat{a}_\mathrm{s} +  \chi_3^{\{\mathrm{p,s,i}\}} \sqrt{\frac{C_\mathrm{g}N}{2\hbar \omega_\mathrm{p}}} |A_{\mathrm{p},0}| \hat{a}_\mathrm{s}^\dagger e^{-i\chi_1^\mathrm{p} t} \right]
\end{align}

in the co-rotating frame
\begin{align}
    \frac{\mathrm{d}\hat{a}_\mathrm{s}}{\mathrm{d}t}
    &= -i\chi_3^{\{\mathrm{p,s,i}\}} \sqrt{\frac{C_\mathrm{g}N}{2\hbar \omega_\mathrm{p}}} |A_{\mathrm{p},0}| \hat{a}_\mathrm{i}^\dagger e^{-i(\chi_1^\mathrm{p} -\chi_1^\mathrm{s} - \chi_1^\mathrm{i})t}\nonumber\\
    &=-i\chi_3 |A_{\mathrm{p},0}| \hat{a}_\mathrm{i}^\dagger e^{-i\Psi_3 t}
\end{align}

\begin{align}
    \frac{\mathrm{d}\hat{a}_\mathrm{i}}{\mathrm{d}t}
    &=-i\chi_3 ^{\{\mathrm{p,s,i}\}}\sqrt{\frac{C_\mathrm{g}N}{2\hbar \omega_\mathrm{p}}} |A_{\mathrm{p},0}| \hat{a}_\mathrm{s}^\dagger e^{-i(\chi_1^\mathrm{p} -\chi_1^\mathrm{s} - \chi_1^\mathrm{i})t}\nonumber\\ 
    &=-i\chi_3 |A_{\mathrm{p},0}| \hat{a}_\mathrm{s}^\dagger e^{-i\Psi_3 t}
\end{align}

with
\begin{align}\label{eq:phase_mismatch_density_3WM}
    \Psi_3 = \chi_1^\mathrm{p} -\chi_1^\mathrm{s} - \chi_1^\mathrm{i}
\end{align}

and
\begin{align}\label{eq_chi3}
    \chi_3=\sqrt{\frac{C_\mathrm{g}N}{2\hbar \omega_\mathrm{p}}}\chi_3^{\{\mathrm{p,s,i}\}}
\end{align}

\section{Squeezing}\label{appendix:squeezing}

The correlation of the signal and idler photons results in a so-called squeezed output field of a JTWPA. One can define the thermal photon number as
\begin{align}\label{eq:N_ThermalPhotonNumber}
    N(\omega)=\sum_n\Big(\braket{ \hat{a}_{\omega}^{\dagger}  \hat{a}_{\omega_n} }
    - \braket{ \hat{a}_{\omega}^{\dagger}}\braket{ \hat{a}_{\omega_n}}\Big)
\end{align}

and the squeezing parameter as
\begin{align}\label{eq:M_SqueezingParameter}
    M(\omega)=\sum_n\Big(\braket{ \hat{a}_{\omega}  \hat{a}_{\omega_n} }
    - \braket{ \hat{a}_{\omega}}\braket{ \hat{a}_{\omega_n}}\Big)
\end{align}

From the definitions \eqref{eq:Y_Quadratures} and \eqref{quadrature_fluctuation} one can compute the relation between the squeezing spectrum ($S$), the thermal photon number and the squeezing parameter
\begin{align}
    S(\omega)&=\sum_n\braket{\Delta\hat{Y}^{\theta}(\omega) \Delta\hat{Y}^{\theta}(\omega_n)}\nonumber\\
    &=1 + 2N(\omega)-2|M(\omega)|
\end{align}

For a vacuum input state, the number of thermal photons can be easily calculated through equations \eqref{eq49}
\begin{align}\label{eq:N_VacInput}
    N(\omega)=|v(\omega,t)|^2
\end{align}

Again using \eqref{eq49}, the squeezing parameter for a vacuum input state can be written as
\begin{align}\label{eq:M_VacInput}
    M(\omega)&= \sum_n\Big(\braket{ \hat{a}_{\omega}  \hat{a}_{\Omega_n} }
    - \braket{ \hat{a}_{\omega}}\braket{ \hat{a}_{\Omega_n}}\Big) \nonumber\\
    &= \sum_n\braket{ \hat{a}_{\omega}  \hat{a}_{\Omega_n } } \nonumber\\
    &= \sum_n \bra{vac}\Big(u(\omega, t)\hat{a}_{\omega,0} +  iv(\omega, t)\hat{a}_{\omega',0}^{\dagger}\Big)\cdot\nonumber\\
    & \hspace{0.4cm}\cdot \Big(u(\Omega_n, t)\hat{a}_{\Omega_{n,0}} + iv(\Omega_n, t)\hat{a}_{\Omega'_{n,0}}^{\dagger}\Big)\ket{vac}e^{-i\Psi t} \nonumber\\
    &= iu(\omega,t) e^{-i\Psi t} \sum_n v(\Omega_n,t)\bra{vac}\hat{a}_{\omega,0}\Big(\hat{a}_{\Omega_n,0}\Big)^{\dagger} \ket{vac} \nonumber\\
    &= iu(\omega,t)v(\omega,t)e^{-i\Psi t} \nonumber\\
    &= \Big(\frac{\Psi \chi_3|A_{\mathrm{p},0}|}{2g^2}\sinh^2{(gt)} -i\frac{\chi_3|A_{\mathrm{p},0}|}{g}\sinh{(gt)}\cosh{(gt)} \Big)e^{-i\Psi t} \nonumber\\
    &= |u(\omega,t)v(\omega,t)|e^{-i \Big( \arctan{\Big(\frac{2g}{\Psi}\coth{(gt)}\Big)} + \Psi t\Big)} = \nonumber\\
    &=|M(\omega)|e^{i\theta}
\end{align}

Where we exploited $v(\omega')=v(\omega)$ and identified the squeezing angle as
\begin{align}\label{eq:theta}
    \theta&= - \Big( \arctan{\Big(\frac{2g}{\Psi}\coth{gt}\Big)} + \Psi t\Big)
\end{align} 

Hence one can easily find the relation between $M(\omega)$ and $N(\omega)$ as
\begin{align}\label{eq:SQL} 
    |M(\omega)|^2&= |u(\omega,t)v(\omega,t)|^2 \nonumber\\
    &= |u(\omega,t)|^2|v(\omega,t)|^2 \nonumber\\
    &=\Big(|v(\omega,t)|^2 + 1\Big)|v(\omega,t)|^2 \nonumber\\
    &=  N(\omega)[N(\omega) + 1]
\end{align}

that is the maximum allowed by the Heisenberg uncertainty principle and implies that the amplification is quantum limited \cite{PhysRevLett.112.190504}.\\

\end{document}